\documentclass[num-refs,sort&compress]{wiley-article}
\usepackage{graphicx,dcolumn,bm,xcolor,microtype,multirow,amscd,amsmath,amssymb,amsfonts,physics,longtable,mhchem,siunitx,rotating,threeparttable,threeparttablex,ntheorem}

\usepackage{soul}

\usepackage[
	colorlinks=true,
    citecolor=blue,
    breaklinks=true
	]{hyperref}
\newcommand{\ra}{\rightarrow}
\newcommand{\pis}{\pi^*}

\newcommand{\ie}{\textit{i.e.}}
\newcommand{\eg}{\textit{e.g.}}
\newcommand{\alert}[1]{\textcolor{black}{#1}}
\newcommand{\mc}{\multicolumn}

\newcommand{\SupInf}{supporting information}

\papertype{Review Article}
\paperfield{Journal Section}

\title{QUESTDB: a database of highly-accurate excitation energies for the electronic structure community}

\author[1]{Micka\"el V\'eril}
\author[1]{Anthony Scemama}
\author[1]{Michel Caffarel}
\author[2]{Filippo Lipparini}
\author[1]{Martial Boggio-Pasqua}
\author[3]{Denis Jacquemin}
\author[1]{Pierre-Fran\c{c}ois Loos}

\affil[1]{Laboratoire de Chimie et Physique Quantiques, Universit\'e de Toulouse, CNRS, UPS, France}
\affil[2]{Dipartimento di Chimica e Chimica Industriale, University of Pisa, Via Moruzzi 3, 56124 Pisa, Italy}
\affil[3]{Universit\'e de Nantes, CNRS,  CEISAM UMR 6230, F-44000 Nantes, France}

\corraddress{Denis Jacquemin and Pierre-Fran\c{c}ois Loos}
\corremail{denis.jacquemin@univ-nantes.fr; loos@irsamc-ups-tlse.fr}

\fundinginfo{European Research Council (ERC), European Union's Horizon 2020 research and innovation programme, Grant agreement No.~863481.
Agence Nationale de la Recherche (ANR) in the framework of the PIA program ANR-18-EURE-0012.
}

\runningauthor{V\'eril et al.}

\begin{document}

\maketitle

\begin{abstract}
We describe our efforts of the past few years to create a large set of more than 500 highly-accurate vertical excitation energies of various natures ($\pi \to \pis$, $n \to \pis$, double excitation,
Rydberg, singlet, doublet, triplet, etc) in small- and medium-sized molecules. These values have been obtained using an incremental strategy which consists in combining high-order coupled
cluster and selected configuration interaction calculations using increasingly large diffuse basis sets in order to reach high accuracy. One of the key aspect of the so-called QUEST database
of vertical excitations is that it does not rely on any experimental values, avoiding potential biases inherently linked to experiments and facilitating  theoretical cross comparisons. Following this
composite protocol, we have been able to produce theoretical best estimate (TBEs) with the aug-cc-pVTZ basis set for each of these transitions, as well as basis set corrected TBEs (i.e., near
the complete basis set limit) for some of them. The TBEs/aug-cc-pVTZ have been employed to benchmark a large number of (lower-order) wave function methods such as CIS(D), ADC(2), CC2,
STEOM-CCSD, CCSD, CCSDR(3), CCSDT-3, ADC(3), CC3,  NEVPT2, and others (including spin-scaled variants). In order to gather the huge amount of data produced during the QUEST
project, we have created a website [\url{https://lcpq.github.io/QUESTDB_website}] where one can easily test and compare the accuracy of a given method with respect to various variables
such as the molecule size or its family,  the nature of the excited states, the type of basis set, etc.
We hope that the present review will provide a useful summary of our effort so far and foster new developments around excited-state methods.
% Please include a maximum of seven keywords
\keywords{excited states, benchmark, database, full configuration interaction, coupled cluster theory, excitation energies}
\end{abstract}

%%%%%%%%%%%%%%%%%%%%%%%%%%%%%
\section{Introduction}
%%%%%%%%%%%%%%%%%%%%%%%%%%%%%

Nowadays, there exist a very large number of electronic structure computational approaches, more or less expensive depending on their overall accuracy, able to quantitatively predict the
absolute and/or relative energies of electronic states in molecular systems \cite{SzaboBook,JensenBook,CramerBook,HelgakerBook}. One important aspect of some of these theoretical
methods is their ability to access the energies of electronic excited states, i.e., states that have higher total energies than the so-called ground (that is, lowest-energy) state
\cite{Roos_1996,Piecuch_2002,Dreuw_2005,Krylov_2006,Sneskov_2012,Gonzales_2012,Laurent_2013,Adamo_2013,Ghosh_2018,Blase_2020,Loos_2020a}.
The faithful description of excited states is particularly challenging from a theoretical point of view but is key to a deeper understanding of photochemical and photophysical processes
like absorption, fluorescence, phosphorescence, chemoluminescence, and others \cite{Bernardi_1996,Olivucci_2010,Robb_2007,Navizet_2011,Crespo_2018,Robb_2018,Mai_2020}.
For a given level of theory, ground-state methods are usually more accurate than their excited-state analogs.
The reasons behind this are (at least) threefold:
i) accurately modeling the electronic structure of excited states usually requires larger one-electron basis sets (including diffuse functions most of the times) than their ground-state counterpart,
ii) excited states can be governed by different amounts of dynamic/static correlations, present very different physical natures ($\pi \to \pis$, $n \to \pis$, charge transfer, double excitation, valence, Rydberg, singlet, doublet, triplet, etc), yet be very close in energy from one another, and
iii) one usually has to rely on response theory formalisms \cite{Monkhorst_1977,Helgaker_1989,Koch_1990,Koch_1990b,Christiansen_1995b,Christiansen_1998b,Hattig_2003,Kallay_2004,Hattig_2005c}, which inherently introduce a ground-state ``bias''.
Hence, designing excited-state methods able to tackle simultaneously and on an equal footing all these types of excited states at an affordable cost remains an open challenge in theoretical computational chemistry as evidenced by the large number of review
 articles on this particular subject \cite{Roos_1996,Piecuch_2002,Dreuw_2005,Krylov_2006,Sneskov_2012,Gonzales_2012,Laurent_2013,Adamo_2013,Dreuw_2015,Ghosh_2018,Blase_2020,Loos_2020a}.

When designing a new theoretical model, the first feature that one might want to test is its overall accuracy, i.e., its ability to reproduce reference (or benchmark) values for a given system with a well-defined
setup (same geometry, basis set, etc). These values can be absolute and/or relative energies, geometrical parameters, physical or chemical spectroscopic properties extracted from experiments,
high-level theoretical calculations, or any combination of these. To this end, the electronic structure community has designed along the years benchmark sets, i.e., sets of molecules for which one
can (very) accurately compute theoretical estimates and/or access solid experimental data for given properties. Regarding ground-states properties, two of the oldest and most employed sets are
probably the Gaussian-1 and Gaussian-2 benchmark sets \cite{Pople_1989,Curtiss_1991,Curtiss_1997} developed by the group of Pople in the 1990's. For example, the Gaussian-2 set gathers atomization
energies, ionization energies, electron affinities, proton affinities, bond dissociation energies, and reaction barriers. This set was subsequently extended and refined \cite{Curtiss_1998,Curtiss_2007}.
Another very useful set for the design of methods able to catch dispersion effects \cite{Angyan_2020} is the S22 benchmark set \cite{Jureka_2006} (and its extended S66 version \cite{Rezac_2011})
of Hobza and collaborators which provides benchmark interaction energies for weakly-interacting (non covalent) systems. One could also mentioned the $GW$100 set \cite{vanSetten_2015,Krause_2015,Maggio_2016}
(and its $GW$5000 extension \cite{Stuke_2020}) of ionization energies which has helped enormously the community to compare the implementation of $GW$-type methods for molecular
systems \cite{vanSetten_2013,Bruneval_2016,Caruso_2016,Govoni_2018}. The extrapolated ab initio thermochemistry (HEAT) set designed to achieve high accuracy for enthalpies of formation
of atoms and small molecules (without experimental data) is yet another successful example of benchmark set \cite{Tajti_2004,Bomble_2006,Harding_2008}. More recently, let us mention the benchmark datasets
of the \textit{Simons Collaboration on the Many-Electron Problem} providing, for example, highly-accurate ground-state energies for
hydrogen chains \cite{Motta_2017} as well as transition metal atoms and their ions and monoxides \cite{Williams_2020}. 
Let us also mention the set of Zhao and Truhlar for small transition metal complexes employed to compare the accuracy of density-functional methods \cite{ParrBook} for $3d$ transition-metal chemistry \cite{Zhao_2006}, \alert{the MGCDB84 molecular database of Mardirossian and Head-Gordon that they used to benchmark a total of 200 density functionals and design (using a combinatorial approach) the $\omega$B97M-V functional \cite{Mardirossian_2017},} and finally the popular GMTKN24 \cite{Goerigk_2010},
GMTKN30 \cite{Goerigk_2011a,Goerigk_2011b} and GMTKN55 \cite{Goerigk_2017} databases for general main group thermochemistry, kinetics, and non-covalent interactions developed by Goerigk, Grimme and
their coworkers.

The examples of benchmark sets presented above are all designed for ground-state properties, and there  exists specific protocols taylored to accurately model excited-state energies and properties as well.
Indeed, benchmark datasets of excited-state energies and/or properties are less numerous than their ground-state counterparts but their number has been growing at a consistent pace in the past few years.
Below, we provide a short description for some of them. One of the most characteristic example is the benchmark set of vertical excitation energies proposed by Thiel and coworkers
\cite{Schreiber_2008,Silva-Junior_2008,Silva-Junior_2010,Silva-Junior_2010b,Silva-Junior_2010c}. The so-called Thiel (or M\"ulheim) set of excitation energies gathers a large number of excitation energies
determined in 28 medium-sized organic CNOH molecules with a total of 223 valence excited states (152 singlet and 71 triplet states) for which theoretical best estimates (TBEs) were defined.
In their first study, Thiel and collaborators performed CC2 \cite{Christiansen_1995a,Hattig_2000}, CCSD \cite{Rowe_1968,Koch_1990,Stanton_1993,Koch_1994}, CC3 \cite{Christiansen_1995b,Koch_1997}, and
CASPT2 \cite{Andersson_1990,Andersson_1992,Roos,Roos_1996} calculations (with the TZVP basis) on MP2/6-31G(d) geometries in order to provide (based on additional high-quality literature data) TBEs for these
transitions. These TBEs were quickly refined with the larger aug-cc-pVTZ basis set \cite{Silva-Junior_2010b,Silva-Junior_2010c}. In the same spirit, it is also worth mentioning Gordon's set of vertical transitions
(based on experimental values) \cite{Leang_2012} used to benchmark the performance of time-dependent density-functional theory (TD-DFT) \cite{Runge_1984,Casida_1995,Casida_2012,Ulrich_2012}, as well
as its extended version by Goerigk and coworkers who decided to replace the experimental reference values by CC3 excitation energies \cite{Schwabe_2017,Casanova-Paez_2019,Casanova_Paes_2020}.
For comparisons with experimental values, there also exists various sets of measured 0-0 energies used in various benchmarks, notably by the Furche \cite{Furche_2002,Send_2011a}, H\"attig \cite{Winter_2013}
and our  \cite{Loos_2018,Loos_2019a,Loos_2019b}  groups for gas-phase compounds and by Grimme \cite{Dierksen_2004,Goerigk_2010a} and one of us \cite{Jacquemin_2012,Jacquemin_2015b} for solvated dyes.
Let us also mention the new benchmark set of charge-transfer excited states recently introduced by Szalay and coworkers [based on equation-of-motion coupled cluster (EOM-CC) methods] \cite{Kozma_2020}
as well as the Gagliardi-Truhlar set employed to compare the accuracy of multiconfiguration pair-density functional theory \cite{Ghosh_2018} against the well-established CASPT2 method \cite{Hoyer_2016}.

\begin{figure}
	\centering
	\includegraphics[width=0.6\linewidth]{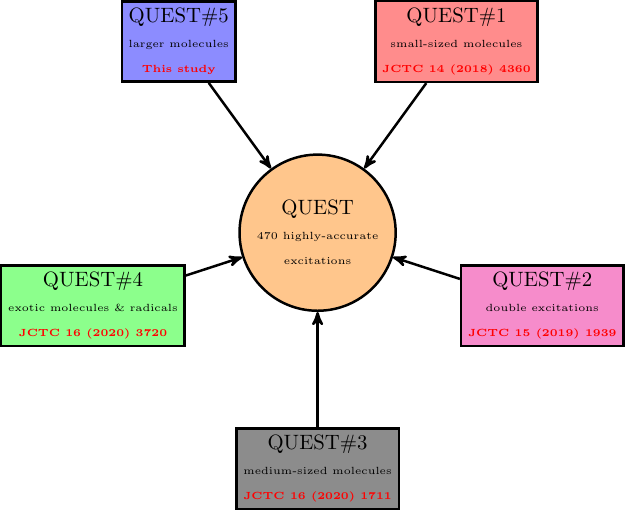}
	\caption{Composition of each of the five subsets making up the present QUEST dataset of highly-accurate vertical excitation energies.}
	\label{fig:scheme}
\end{figure}

Following a similar philosophy and striving for chemical accuracy, we have recently reported in several studies highly-accurate vertical excitations for small- and medium-sized molecules
\cite{Loos_2020a,Loos_2018a,Loos_2019,Loos_2020b,Loos_2020c}. The so-called QUEST dataset of vertical excitations which we will describe in detail in the present review article is composed by 5
subsets (see Fig.~\ref{fig:scheme}): i) a subset of excitations in small molecules containing from 1 to 3 non-hydrogen atoms known as QUEST\#1, ii) a subset of double excitations in molecules of small and
medium sizes known as QUEST\#2, iii) a subset of excitation energies for medium-sized molecules containing from 4 to 6 non-hydrogen atoms known as QUEST\#3, iv) a subset composed by more ``exotic''
molecules and radicals labeled as QUEST\#4, and v) a subset known as QUEST\#5, specifically designed for the present article, gathering excitation energies in larger molecules as well as additional smaller molecules.
One of the key aspect of the QUEST dataset is that it does not rely on any experimental values, avoiding potential biases inherently linked to experiments and facilitating in the process theoretical comparisons.
Moreover, our protocol has been designed to be as uniform as possible, which means that we have designed a very systematic procedure for all excited states in order to make cross-comparison as straightforward as possible.
Importantly, it allowed us to benchmark, in a very systematic and balanced way, a series of popular excited-state wave function methods partially or fully accounting for double and triple excitations as well as multiconfigurational methods (see below).
In the same vein, as evoked above, we have also produced chemically-accurate theoretical 0-0 energies \cite{Loos_2018,Loos_2019a,Loos_2019b} which can be more straightforwardly compared to experimental data \cite{Furche_2002,Kohn_2003,Dierksen_2004,Goerigk_2010a,Send_2011a,Jacquemin_2012,Winter_2013,Fang_2014,Jacquemin_2015b,Oruganti_2016}. We refer the interested reader to Ref.~\cite{Loos_2019b} for a
review of the generic benchmark studies devoted to adiabatic and 0-0 energies performed in the past two decades.

%%% FIGURE 1 %%%

The QUEST dataset has the particularity to be based to a large extent on selected configuration interaction (SCI) reference excitation energies as well as high-order linear-response (LR) CC methods such as LR-CCSDT and
LR-CCSDTQ \cite{Noga_1987,Koch_1990,Kucharski_1991,Christiansen_1998b,Kucharski_2001,Kowalski_2001,Kallay_2003,Kallay_2004,Hirata_2000,Hirata_2004}. Recently, SCI methods have been a force to reckon with for
the computation of highly-accurate energies in small- and medium-sized molecules as they yield near full configuration interaction (FCI) quality energies for only a very tiny fraction of the computational cost of a genuine FCI calculation \cite{Booth_2009,Booth_2010,Cleland_2010,Booth_2011,Daday_2012,Blunt_2015,Ghanem_2019,Deustua_2017,Deustua_2018,Holmes_2017,Chien_2018,Li_2018,Yao_2020,Li_2020,Eriksen_2017,Eriksen_2018,Eriksen_2019a,Eriksen_2019b,Xu_2018,Xu_2020,Loos_2018a,Loos_2019,Loos_2020b,Loos_2020c,Loos_2020a,Loos_2020e,Eriksen_2021}.
Due to the fairly natural idea underlying these methods, the SCI family is composed of numerous members \cite{Bender_1969,Whitten_1969,Huron_1973,Abrams_2005,Bunge_2006,Bytautas_2009,Giner_2013,Caffarel_2014,Giner_2015,Garniron_2017b,Caffarel_2016a,Caffarel_2016b,Holmes_2016,Sharma_2017,Holmes_2017,Chien_2018,Scemama_2018,Scemama_2018b,Garniron_2018,Evangelista_2014,Tubman_2016,Tubman_2020,Schriber_2016,Schriber_2017,Liu_2016,Per_2017,Ohtsuka_2017,Zimmerman_2017,Li_2018,Ohtsuka_2017,Coe_2018,Loos_2019}.
Their fundamental philosophy consists, roughly speaking, in retaining only the most relevant determinants of the FCI space following a given criterion to slow down the exponential increase of the size of the CI expansion.
Originally developed in the late 1960's by Bender and Davidson \cite{Bender_1969} as well as Whitten and Hackmeyer \cite{Whitten_1969}, new efficient SCI algorithms have resurfaced recently.
Three examples are iCI \cite{Liu_2014,Liu_2016,Lei_2017,Zhang_2020}, semistochastic heat-bath CI (SHCI) \cite{Holmes_2016,Holmes_2017,Sharma_2017,Li_2018,Li_2020,Yao_2020}, and \textit{Configuration Interaction using a Perturbative Selection made Iteratively} (CIPSI) \cite{Huron_1973,Giner_2013,Giner_2015,Garniron_2019}.
These flavors of SCI include a second-order perturbative (PT2) correction which is key to estimate the ``distance'' to the FCI solution (see below).
The SCI calculations performed for the QUEST set of excitation energies relies on the CIPSI algorithm, which is, from a historical point of view, one of the oldest SCI algorithms.
It was developed in 1973 by Huron, Rancurel, and Malrieu \cite{Huron_1973} (see also Refs.~\cite{Evangelisti_1983,Cimiraglia_1985,Cimiraglia_1987,Illas_1988,Povill_1992}).
Recently, the determinant-driven CIPSI algorithm has been efficiently implemented \cite{Garniron_2019} in the open-source programming environment QUANTUM PACKAGE by the Toulouse group enabling to perform massively
parallel computations \cite{Garniron_2017,Garniron_2018,Garniron_2019,Loos_2020e}. CIPSI is also frequently employed to provide accurate trial wave functions for quantum Monte Carlo calculations in molecules \cite{Caffarel_2014,Caffarel_2016a,Caffarel_2016b,Giner_2013,Giner_2015,Scemama_2015,Scemama_2016,Scemama_2018,Scemama_2018b,Scemama_2019,Dash_2018,Dash_2019,Scemama_2020} and more recently
for periodic solids \cite{Benali_2020}. We refer the interested reader to Ref.~\cite{Garniron_2019} where one can find additional details regarding the implementation of the CIPSI algorithm.

The present article is organized as follows. In Sec.~\ref{sec:tools}, we detail the specificities of our protocol by providing computational details regarding geometries, basis sets, (reference and benchmarked)
computational methods, and a new way of estimating rigorously the extrapolation error in SCI calculations which is tested by computing additional FCI values for five- and six-membered rings.
We then describe in Sec.~\ref{sec:QUEST} the content of our five QUEST subsets providing for each of them the number of reference excitation energies, the nature and size of the molecules, the list of
benchmarked methods, as well as other specificities. A special emphasis is placed on our latest (previously unpublished) add-on, QUEST\#5, specifically designed for the present manuscript where we have considered, in particular
but not only, larger molecules. Section \ref{sec:TBE} discusses the generation of the TBEs, while Sec.~\ref{sec:bench} proposes a comprehensive benchmark of various methods on the entire QUEST set which is
composed by more than 400 excitations with, in addition, a specific analysis for each type of excited states. Section \ref{sec:website} describes the feature of the website that we have specifically designed to gather the
entire data generated during these last few years. Thanks to this website, one can easily test and compare the accuracy of a given method with respect to various variables such as the molecule size or its family, the nature
of the excited states, the size of the basis set, etc. Finally, we draw our conclusions in Sec.~\ref{sec:ccl} where we discuss, in particular, future projects aiming at expanding and improving the usability and accuracy of the QUEST database.

%%%%%%%%%%%%%%%%%%%%%%%%%%%%%
\section{Computational tools}
\label{sec:tools}
%%%%%%%%%%%%%%%%%%%%%%%%%%%%%

%=======================
\subsection{Geometries}
%=======================
The ground-state structures of the molecules included in the QUEST dataset have been systematically optimized at the CC3/aug-cc-pVTZ level of theory, except for a very few cases.
As shown in Refs.~\cite{Hattig_2005c,Budzak_2017}, CC3 provides extremely accurate ground- and excited-state geometries. These optimizations have been performed using DALTON 2017
\cite{dalton} and CFOUR 2.1 \cite{cfour} applying default parameters.  For the open-shell derivatives belonging to QUEST\#4 \cite{Loos_2020c}, the geometries are optimized at the UCCSD(T)/aug-cc-pVTZ level using the GAUSSIAN16 program \cite{Gaussian16} and applying the ``tight'' convergence threshold.  For the purpose of the present review article, we have gathered all the geometries in the {\SupInf}.

%=======================
\subsection{Basis sets}
%=======================
For the entire set, we rely on the 6-31+G(d) Pople basis set \cite{Binkley_1977a,Clark_1983a,Dill_1975a,Ditchfield_1971a,Francl_1982a,Gordon_1982a,Hehre_1972a}, the augmented family of Dunning basis sets aug-cc-pVXZ (where X $=$ D, T, Q, and 5) \cite{Dunning_1989a,Kendall_1992a,Prascher_2011a,Woon_1993a,Woon_1994a}, and sometimes its doubly- and triply-augmented variants, d-aug-cc-pVXZ and t-aug-cc-pVXZ respectively.
Doubly- and triply-augmented basis sets are usually employed for Rydberg states where it is not uncommon to observe a strong basis set dependence due to the very diffuse nature of these excited states.
These basis sets are available from the \href{https://www.basissetexchange.org}{basis set exchange} website \cite{Feller_1996a,Pritchard_2019a,Schuchardt_2007a}.

%==================================
\subsection{Computational methods}
%==================================
\label{sec:methods}
%------------------------------------------------
\subsubsection{Reference computational methods}
%------------------------------------------------
In order to compute reference vertical energies, we have designed different strategies depending on the actual nature of the transition and the size of the system.
For small molecules (typically 1--3 non-hydrogen atoms), we mainly resort to SCI methods which can provide near-FCI excitation energies for compact basis sets.
Obviously, the smaller the molecule, the larger the basis we can afford.
For larger systems (\ie, 4--6 non-hydrogen atom), one cannot afford SCI calculations anymore except in a few special occasions, and we then rely on LR-CC theory (LR-CCSDT and LR-CCSDTQ typically \cite{Kucharski_1991,Kallay_2003,Kallay_2004,Hirata_2000,Hirata_2004}) to obtain accurate transition energies.
In the following, we will omit the prefix LR for the sake of clarity, as equivalent values would be obtained with the equation-of-motion (EOM) formalism \cite{Rowe_1968,Stanton_1993}.

The CC calculations are performed with several codes.
For closed-shell molecules, CC3 \cite{Christiansen_1995b,Koch_1997} calculations are achieved with DALTON \cite{dalton} and CFOUR \cite{cfour}.
CCSDT and CCSDTQ calculations are performed with CFOUR \cite{cfour} and MRCC 2017 \cite{Rolik_2013,mrcc}, the latter code being also used for  CCSDTQP.
%Note that all our excited-state CC calculations are performed within the equation-of-motion (EOM) or linear-response (LR) formalism that yield the same excited-state energies.
The reported oscillator strengths have been computed in the LR-CC3 formalism only.
For open-shell molecules, the CCSDT, CCSDTQ, and CCSDTQP calculations performed with MRCC \cite{Rolik_2013,mrcc} do consider an unrestricted Hartree-Fock wave function as reference but for a few exceptions.
All excited-state calculations are performed, except when explicitly mentioned, in the frozen-core (FC) approximation using large cores for the third-row atoms.

All the SCI calculations are performed within the frozen-core approximation using QUANTUM PACKAGE \cite{Garniron_2019} where the CIPSI algorithm \cite{Huron_1973} is implemented. Details regarding this specific CIPSI implementation can be found in Refs.~\cite{Garniron_2019} and \cite{Scemama_2019}.
A state-averaged formalism is employed, i.e., the ground and excited states are described with the same set of determinants and orbitals, but different CI coefficients.
Our usual protocol \cite{Scemama_2018,Scemama_2018b,Scemama_2019,Loos_2018a,Loos_2019,Loos_2020a,Loos_2020b,Loos_2020c} consists of performing a preliminary CIPSI calculation using Hartree-Fock orbitals in order to generate a CIPSI wave function with at least $10^7$ determinants.
Natural orbitals are then computed based on this wave function, and a new, larger CIPSI calculation is performed with this new set of orbitals.
This has the advantage to produce a smoother and faster convergence of the SCI energy toward the FCI limit.
The CIPSI energy $E_\text{CIPSI}$ is defined as the sum of the variational energy $E_\text{var}$ (computed via diagonalization of the CI matrix in the reference space) and a PT2 correction $E_\text{PT2}$ which estimates the contribution of the determinants not included in the CI space \cite{Garniron_2017b}.
By linearly extrapolating this second-order correction to zero, one can efficiently estimate the FCI limit for the total energies.
These extrapolated total energies (simply labeled as $E_\text{FCI}$ in the remainder of the paper) are then used to compute vertical excitation energies.

Depending on the set, we estimated the extrapolation error via different techniques.
For example, in Ref.~\cite{Loos_2020b}, we estimated the extrapolation error by the difference between the transition energies obtained with the largest SCI wave function and the FCI extrapolated value.
This definitely cannot be viewed as a true error bar, but it provides an idea of the quality of the FCI extrapolation and estimate.
Below, we provide a much cleaner way of estimating the extrapolation error in SCI methods, and we adopt this scheme for the five- and six-membered rings considered in the QUEST\#3 subset.
The particularity of the current implementation is that the selection step and the PT2 correction are computed \textit{simultaneously} via a hybrid semistochastic algorithm \cite{Garniron_2017,Garniron_2019}.
Moreover, a renormalized version of the PT2 correction (dubbed rPT2) has been recently implemented for a more efficient extrapolation to the FCI limit \cite{Garniron_2019}.
We refer the interested reader to Ref.~\cite{Garniron_2019} where one can find all the details regarding the implementation of the CIPSI algorithm.
\alert{Note that all our SCI wave functions are eigenfunctions of the $\Hat{S}^2$ spin operator which is, unlike ground-state calculations, paramount in the case of excited states \cite{Applencourt_2018}. In the case of ground-state calculations, this constraint can be relaxed without altering the final result (see, for example, Ref.~\cite{Loos_2020e}).}

%------------------------------------------------
\subsubsection{Benchmarked computational methods}
%------------------------------------------------

Using a large variety of codes, our benchmark effort consists in evaluating the accuracy of vertical transition energies obtained at lower levels of theory.
For example, we rely on GAUSSIAN \cite{Gaussian16} and TURBOMOLE 7.3 \cite{Turbomole} for CIS(D) \cite{Head-Gordon_1994,Head-Gordon_1995};
Q-CHEM 5.2 \cite{Krylov_2013} for EOM-MP2 [CCSD(2)] \cite{Stanton_1995c} and ADC(3) \cite{Trofimov_2002,Harbach_2014,Dreuw_2015};
Q-CHEM \cite{Krylov_2013} and TURBOMOLE \cite{Turbomole} for ADC(2) \cite{Trofimov_1997,Dreuw_2015};
DALTON \cite{dalton} and TURBOMOLE \cite{Turbomole} for CC2 \cite{Christiansen_1995a,Hattig_2000};
DALTON \cite{dalton} and GAUSSIAN \cite{Gaussian16} for CCSD \cite{Koch_1990,Stanton_1993,Koch_1994};
DALTON \cite{dalton} for CCSDR(3) \cite{Christiansen_1996b};
CFOUR \cite{cfour} for CCSDT-3 \cite{Watts_1996b,Prochnow_2010};
and ORCA \cite{Neese_2012} for similarity-transformed EOM-CCSD (STEOM-CCSD) \cite{Nooijen_1997,Dutta_2018}.
In addition, we evaluate the spin-opposite scaling (SOS) variants of ADC(2), SOS-ADC(2), as implemented in both Q-CHEM \cite{Krauter_2013} and TURBOMOLE \cite{Hellweg_2008}.
Note that these two codes have distinct SOS implementations, as explained in Ref.~\cite{Krauter_2013}.
We also test the SOS and spin-component scaled (SCS) versions of CC2, as implemented in TURBOMOLE \cite{Hellweg_2008,Turbomole}.
Discussion of various spin-scaling schemes can  be found elsewhere \cite{Goerigk_2010a}.
%When available, we take advantage of the resolution-of-the-identity (RI) approximation in TURBOMOLE and Q-CHEM.
For the STEOM-CCSD calculations, it was checked that the active character percentage was, at least, $98\%$.
%When comparisons between various codes/implementations were possible, we could not detect variations in the transition energies larger than $0.01$ eV.
For radicals, we applied both the U (unrestricted) and RO (restricted open-shell) versions of CCSD and CC3 as implemented in the PSI4 code \cite{Psi4} to perform our benchmarks.
Finally, the composite approach, ADC(2.5), which follows the spirit of Grimme's and Hobza's MP2.5 approach \cite{Pitonak_2009} by averaging the ADC(2) and ADC(3) excitation energies, is also tested in the following \cite{Loos_2020d}.

For the double excitations composing the QUEST database, we have performed additional calculations using various multiconfigurational methods.
In particular, state-averaged (SA) CASSCF and CASPT2 \cite{Roos,Andersson_1990} have been performed with MOLPRO (RS2 contraction level) \cite{molpro}.
Concerning the NEVPT2 calculations (which are also performed with MOLPRO), the partially-contracted (PC) and strongly-contracted (SC) variants have been tested \cite{Angeli_2001a,Angeli_2001b,Angeli_2002}.
From a strict theoretical point of view, we point out that PC-NEVPT2 is supposed to be more accurate than SC-NEVPT2 given that it has a larger number of perturbers and greater flexibility.
PC-NEVPT2 calculations were also systematically performed for the QUEST\#3.
In the case of double excitations \cite{Loos_2019}, we have also performed calculations with multi-state (MS) CASPT2 (MS-MR formalism), \cite{Finley_1998} and its extended variant (XMS-CASPT2) \cite{Shiozaki_2011} when there is a strong mixing between states with same spin and spatial symmetries.
The CASPT2 calculations have been performed with level shift and IPEA parameters set to the standard values of $0.3$ and $0.25$ a.u., respectively.
Large active spaces carefully chosen and tailored for the desired transitions have been selected.
The definition of the active space considered for each system as well as the number of states in the state-averaged calculation is provided in their corresponding publication.

%------------------------------------------------
\subsubsection{Estimating the extrapolation error}
\label{sec:error}
%------------------------------------------------
In this section, we present our scheme to estimate the extrapolation error in SCI calculations.
This new protocol is then applied to five- and six-membered ring molecules for which SCI calculations are particularly challenging even for small basis sets.
Note that the present method does only apply to \emph{state-averaged} SCI calculations where ground- and excited-state energies are produced during the same calculation with the same set of molecular orbitals, not to \emph{state-specific} calculations where one computes solely the energy of a single state (like conventional ground-state calculations).

For the $m$th excited state (where $m = 0$ corresponds to the ground state), we usually estimate its FCI energy $E_{\text{FCI}}^{(m)}$ by performing a linear extrapolation of its variational energy $E_\text{var}^{(m)}$ as a function of its rPT2 correction $E_{\text{rPT2}}^{(m)}$ \cite{Holmes_2017, Garniron_2019} using 
\begin{equation}
  E_{\text{var}}^{(m)} \approx E_\text{FCI}^{(m)} - \alpha^{(m)} E_{\text{rPT2}}^{(m)},
  \label{eqx}
\end{equation}
where $E_{\text{var}}^{(m)}$ and $E_{\text{rPT2}}^{(m)}$ are calculated with CIPSI and $E_\text{FCI}^{(m)}$ is the FCI energy to be extrapolated. 
This relation is valid in the regime of a sufficiently large number of determinants where the second-order perturbational correction largely dominates.
In theory, the coefficient $\alpha^{(m)}$ should be equal to one but, in practice, due to the residual higher-order terms, it deviates slightly from unity.

For the largest systems considered here, $\abs{E_{\text{rPT2}}}$ can be as large as 2~eV and, thus,
the accuracy of the excitation energy estimates strongly depends on our ability to compensate the errors in the calculations.
Here, we greatly enhance the compensation of errors by making use of
our selection procedure ensuring that the rPT2 values of both states
match as well as possible (a trick known as PT2 matching
\cite{Dash_2018,Dash_2019}), i.e. $E_{\text{rPT2}}^{(0)} \approx E_{\text{rPT2}}^{(m)}$, and
by using a common set of state-averaged natural orbitals with equal weights for the ground and excited states.
%This last feature tends to make the values of $\alpha^{(0)}$ and $\alpha^{(m)}$ very close to each other, such that the error on the energy difference is decreased.

Using Eq.~\eqref{eqx} the estimated error on the CIPSI energy is calculated as
\begin{equation}
  E_{\text{CIPSI}}^{(m)} - E_{\text{FCI}}^{(m)}
  = \qty(E_\text{var}^{(m)}+E_{\text{rPT2}}^{(m)}) - E_{\text{FCI}}^{(m)}
  = \qty(1-\alpha^{(m)}) E_{\text{rPT2}}^{(m)}
\end{equation}
and thus the extrapolated excitation energy associated with the $m$th excited state is given by
\begin{equation}
 \Delta E_{\text{FCI}}^{(m)}
 = \qty[ E_\text{var}^{(m)} + E_{\text{rPT2}}^{(m)} + \qty(\alpha^{(m)}-1) E_{\text{rPT2}}^{(m)} ]
 - \qty[ E_\text{var}^{(0)} + E_{\text{rPT2}}^{(0)} + \qty(\alpha^{(0)}-1) E_{\text{rPT2}}^{(0)} ].
\end{equation}
The slopes $\alpha^{(m)}$ and $\alpha^{(0)}$ deviating only slightly from the unity, the error in
$\Delta E_{\text{FCI}}^{(m)}$ can be expressed at leading order as $\qty(\alpha^{(m)}-\alpha^{(0)}) {\bar E}_{\text{rPT2}} + \mathcal{O}\qty[{{\bar E}_{\text{rPT2}}^2}]$, where ${\bar E}_{\text{rPT2}}=\qty(E_{\text{rPT2}}^{(m)} +E_{\text{rPT2}}^{(0)})/2$ is the averaged second-order correction.

\begin{figure}
\centering
\includegraphics[width=0.9\linewidth]{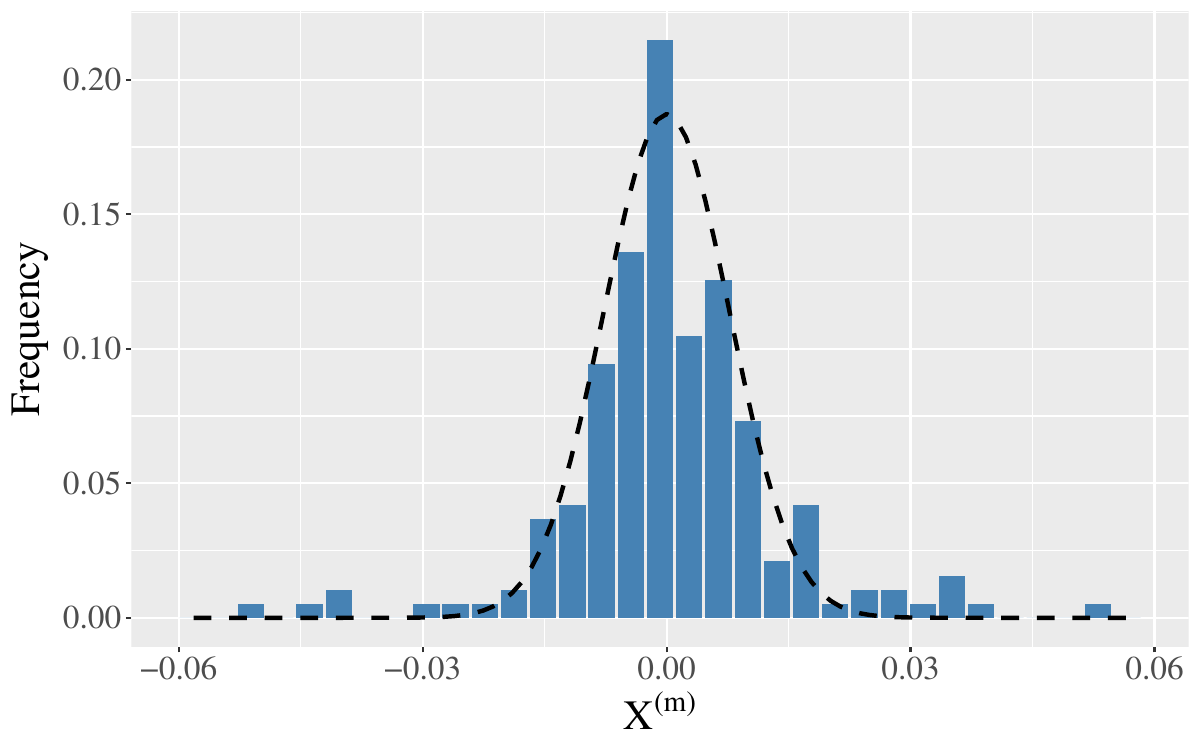}
\caption{Histogram of the random variable $X^{(m)}$ [see Eq.~\eqref{eq:X} in the main text for its definition]. 
About 200 values of singlet and triplet excitation energies taken at various iteration number $n$ for the 13 five- and six-membered ring molecules have been considered to build the present histogram.
The number $M$ of iterations kept at each calculation is chosen according to the statistical test presented in the text.}
\label{fig:histo}
\end{figure}

In the ideal case where one is able to fully correlate the CIPSI calculations associated with the ground and excited states, the fluctuations of
$\Delta E_\text{CIPSI}^{(m)}(n)$ as a function of the iteration number $n$ would completely vanish and the exact excitation energy would be obtained from the first CIPSI iterations.
Quite remarkably, in practice, numerical experience shows that the fluctuations with respect to the extrapolated value $\Delta E_\text{FCI}^{(m)}$ are small,
zero-centered, and display a Gaussian-like distribution.
In addition, as evidenced in Fig.~\ref{fig:histo}, these fluctuations are found to be (very weakly) dependent on the iteration number $n$ (as far as not too close $n$ values are considered). 
Hence, this weak dependency does not significantly alter our results and will not be considered here.

We thus introduce the following random variable
\begin{equation}
\label{eq:X}
	X^{(m)}= \frac{\Delta E_\text{CIPSI}^{(m)}(n)- \Delta E_\text{FCI}^{(m)}}{\sigma(n)}
\end{equation}
where
\begin{equation}
  \Delta E_\text{CIPSI}^{(m)}(n) = \qty[ E_\text{var}^{(m)}(n) +
  E_{\text{rPT2}}^{(m)}(n) ]
  - \qty[ E_\text{var}^{(0)}(n) + E_{\text{rPT2}}^{(0)}(n) ]
\end{equation}
and
${\sigma(n)}$ is a quantity proportional to the average fluctuations of $\Delta E_\text{CIPSI}^{(m)}$.
A natural choice for $\sigma^2(n)$, playing here the role of a variance, is
\begin{equation}
\sigma^2(n) \propto \qty[E_{\text{rPT2}}^{(m)}(n)]^2 + \qty[E_{\text{rPT2}}^{(0)}(n)]^2
\end{equation}
which vanishes in the large-$n$ limit (as it should).

%%% FIGURE 2 %%%

The histogram of $X^{(m)}$ resulting from the singlet and triplet excitation energies obtained at various iteration number $n$ for the 13 five- and six-membered ring molecules is shown in Fig.~\ref{fig:histo}. 
To avoid transient effects, only excitation energies at sufficiently large $n$ are retained in the data set.
The statistical criterion used to decide from which precise value of $n$ the data should be kept is presented below. 
In the present example, the total number of values employed to construct the histogram of Fig.~\ref{fig:histo} is about 200. 
The dashed line represents the best (in a least-squares sense) Gaussian fit reproducing the data.
As clearly seen from Fig.~\ref{fig:histo}, the distribution can be fairly well described by a Gaussian probability distribution
\begin{equation}
	P\qty[X^{(m)}] \propto \exp[-\frac{{X^{(m)}}^2} {2{\sigma^{*}}^2} ]
\end{equation}
where $\sigma^{*2}$ is some ``universal'' variance depending only on the way the correlated selection of both states is done, not on the molecule considered in our set.

For each CIPSI calculation, an estimate of $\Delta E_{\text{FCI}}^{(m)}$ is thus
\begin{equation}
	\Delta E_\text{FCI}^{(m)} = \frac{ \sum_{n=1}^M  \frac{\Delta E_\text{CIPSI}^{(m)}(n)} {\sigma(n)} }
            { \sum_{n=1}^M  \frac{1}{\sigma(n)} }
\end{equation}
where $M$ is the number of iterations that has been retained to compute the statistical quantities.
Regarding the estimate of the error on $\Delta E_\text{FCI}^{(m)}$ some caution is required since, although the distribution is globally Gaussian-like
(see Fig.~\ref{fig:histo}), there exists some significant deviation from it and we must to take this feature into account.

More precisely, we search for a confidence interval $\mathcal{I}$ such that the true value of the excitation energy $\Delta E_{\text{FCI}}^{(m)}$ lies within one standard deviation of $\Delta E_\text{CIPSI}^{(m)}$, i.e., $P\qty( \Delta E_{\text{FCI}}^{(m)} \in \qty[ \Delta E_\text{CIPSI}^{(m)} \pm \sigma ] \; \Big| \; \mathcal{G}) = p = 0.6827$.
In a Bayesian framework, the probability that $\Delta E_{\text{FCI}}^{(m)}$ is in an interval $\mathcal{I}$ is
\begin{equation}
   P\qty( \Delta E_{\text{FCI}}^{(m)} \in \mathcal{I} ) = P\qty( \Delta E_{\text{FCI}}^{(m)} \in I \Big| \mathcal{G}) \times P\qty(\mathcal{G})
\end{equation}
where $P\qty(\mathcal{G})$ is the probability that the random variables considered in the latest CIPSI iterations are normally distributed.
A common test in statistics of the normality of a distribution is the Jarque-Bera test $J$ and we have
\begin{equation}
   P\qty(\mathcal{G}) = 1 - \chi^2_{\text{CDF}}(J,2)
\end{equation}
where $\chi^2_{\text{CDF}}(x,k)$ is the cumulative distribution function (CDF) of the $\chi^2$-distribution with $k$ degrees of freedom.
As the number of samples $M$ is usually small, we use Student's $t$-distribution to estimate the statistical error.
The inverse of the cumulative distribution function of the $t$-distribution, $t_{\text{CDF}}^{-1}$, allows us to find how to scale the interval by a parameter
\begin{equation}
   \beta = t_{\text{CDF}}^{-1} \qty[
   \frac{1}{2} \qty( 1 + \frac{0.6827}{P(\mathcal{G})}), M ]
\end{equation}
such that $P\qty( \Delta E_{\text{FCI}}^{(m)} \in \qty[ \Delta E_{\text{CIPSI}}^{(m)} \pm \beta \sigma ] ) = p $.
Only the last $M>2$ computed transition energies are considered. $M$ is chosen such that $P(\mathcal{G})>0.8$ and such that the error bar is minimal.
If all the values of $P(\mathcal{G})$ are below $0.8$, $M$ is chosen such that $P(\mathcal{G})$ is maximal.
A Python code associated with this procedure is provided in the {\SupInf}.

The singlet and triplet FCI/6-31+G(d) excitation energies and their corresponding error bars estimated with the method presented above based on Gaussian random variables are reported in Table \ref{tab:cycles}.
For the sake of comparison, we also report the CC3 and CCSDT vertical energies from Ref.~\cite{Loos_2020b} computed in the same basis. We note that there is for the vast majority of considered
states a very good agreement between the CC3 and CCSDT values, indicating that the CC values can be trusted.
The estimated values of the excitation energies obtained via a three-point linear extrapolation considering the three largest CIPSI wave functions are also gathered in Table \ref{tab:cycles}.
In this case, the error bar is estimated via the extrapolation distance, \ie, the difference in excitation energies obtained with the three-point linear extrapolation and the largest CIPSI wave function.
This strategy has been considered in some of our previous works \cite{Loos_2020b,Loos_2020c,Loos_2020e}.

\begin{figure}
	\centering
	\includegraphics[width=\linewidth]{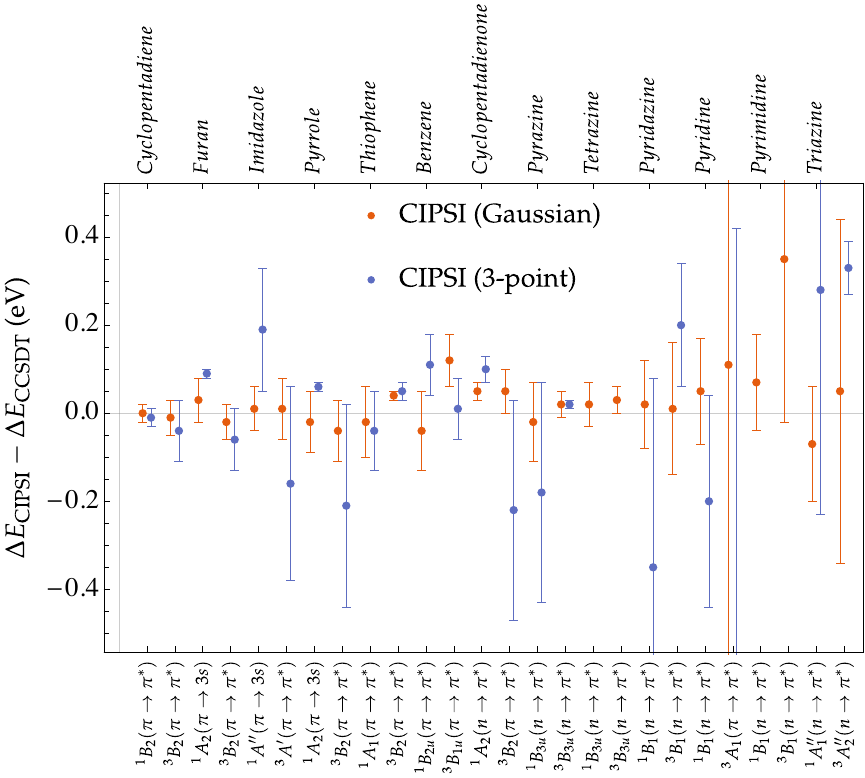}
	\caption{Deviation from the CCSDT excitation energies for the lowest singlet and triplet excitation energies (in eV) of five- and six-membered rings obtained at the CIPSI/6-31+G(d) level of theory. Red dots: excitation energies and error bars estimated via the present method (see Sec.~\ref{sec:error}). Blue dots: excitation energies obtained via a three-point linear fit using the three largest CIPSI wave functions, and error bars estimated via the extrapolation distance, \ie, the difference in excitation energies obtained with the three-point linear extrapolation and the largest CIPSI wave function.}
	\label{fig:errors}
\end{figure}

The deviation from the CCSDT excitation energies for the same set of excitations are depicted in Fig.~\ref{fig:errors}, where the red dots correspond to the excitation energies and error bars estimated via the present method, and the blue dots correspond to the excitation energies obtained via a three-point linear fit and error bars estimated via the extrapolation distance.
These results contain a good balance between well-behaved and ill-behaved cases.
For example, cyclopentadiene and furan correspond to well-behaved scenarios where the two flavors of extrapolations yield nearly identical estimates and the error bars associated with these two methods nicely overlap.
In these cases, one can observe that our method based on Gaussian random variables provides almost systematically smaller error bars.
Even in less idealistic situations (like in imidazole, pyrrole, and thiophene), the results are very satisfactory and stable.
The six-membered rings represent much more challenging cases for SCI methods, and even for these systems the newly-developed method provides realistic error bars, and allows to easily detect problematic events (like pyridine for instance).
The present scheme has also been tested on smaller systems when one can tightly converge the CIPSI calculations.
In such cases, the agreement is nearly perfect in every scenario that we have encountered.
A selection of these results can be found in the {\SupInf}.

%%% TABLE I %%%
\begin{table}
\centering
\caption{Singlet and triplet excitation energies (in eV) obtained at the CC3, CCSDT, and CIPSI levels of theory with the 6-31+G(d) basis set for various five- and six-membered rings.}
\label{tab:cycles}
\begin{threeparttable}
\begin{tabular}{lccccc}
\headrow
\thead{Molecule} 	& \thead{Transition} 				& \thead{CC3}	& \thead{CCSDT} & \thead{CIPSI (Gaussian)$^a$}	& \thead{CIPSI (3-point)$^b$}\\
					\mc{6}{c}{Five-membered rings}	\\
Cyclopentadiene		&	$^1 B_2 (\pi \ra \pis)$			&	5.79	&	5.80	&	5.80(2)	&	5.79(2)		\\
					&	$^3 B_2 (\pi \ra \pis)$			&	3.33	&	3.33	&	3.32(4)	&	3.29(7)		\\
Furan				&	$^1A_2(\pi \ra 3s)$				&	6.26	&	6.28	&	6.31(5)	&	6.37(1)		\\
					&	$^3B_2(\pi \ra \pis)$			&	4.28	&	4.28	&	4.26(4)	&	4.22(7)		\\
Imidazole			&	$^1A''(\pi \ra 3s)$				&	5.77	&	5.77	&	5.78(5)	&	5.96(14)	\\
					&	$^3A'(\pi \ra \pis)$			&	4.83	&	4.81	&	4.82(7)	&	4.65(22)	\\
Pyrrole				&	$^1A_2(\pi \ra 3s)$				&	5.25	&	5.25	&	5.23(7)	&	5.31(1)		\\
					&	$^3B_2(\pi \ra \pis)$			&	4.59	&	4.58	&	4.54(7)	&	4.37(23)	\\
Thiophene			&	$^1A_1(\pi \ra \pis)$			&	5.79	&	5.77	&	5.75(8)	&	5.73(9)		\\
					&	$^3B_2(\pi \ra \pis)$			&	3.95	&	3.94	&	3.98(1)	&	3.99(2)		\\
					\mc{6}{c}{Six-membered rings}	\\
Benzene				&	$^1B_{2u}(\pi \ra \pis)$		&	5.13	&	5.10	&	5.06(9)	&	5.21(7)		\\
					&	$^3B_{1u}(\pi \ra \pis)$		&	4.18	&	4.16	&	4.28(6)	&	4.17(7)		\\
Cyclopentadienone	&	$^1A_2(n \ra \pis)$				&	3.03	&	3.03	&	3.08(2)	&	3.13(3)		\\
					&	$^3B_2(\pi \ra \pis)$			&	2.30	&	2.32	&	2.37(5)	&	2.10(25)	\\
Pyrazine			&	$^1B_{3u}(n \ra \pis)$			&	4.28	&	4.28	&	4.26(9)	&	4.10(25)	\\
					&	$^3B_{3u}(n \ra \pis)$			&	3.68	&	3.68	&	3.70(3)	&	3.70(1)		\\
Tetrazine			&	$^1B_{3u}(n \ra \pis)$			&	2.53	&	2.54	&	2.56(5)	&	5.07(16)	\\
					&	$^3B_{3u}(n \ra \pis)$			&	1.87	&	1.88	&	1.91(3)	&	4.04(49)	\\
Pyridazine			&	$^1B_1(n \ra \pis)$				&	3.95	&	3.95	&	3.97(10)&	3.60(43) 	\\
					&	$^3B_1(n \ra \pis)$				&	3.27	&	3.26	&	3.27(15)&	3.46(14)	\\
Pyridine			&	$^1B_1(n \ra \pis)$				&	5.12	&	5.10	&	5.15(12)&	4.90(24)	\\
					&	$^3A_1(\pi \ra \pis)$			&	4.33	&	4.31	&	4.42(85)&	3.68(105)	\\
Pyrimidine			&	$^1B_1(n \ra \pis)$				&	4.58	&	4.57	&	4.64(11)&	2.54(5)		\\
					&	$^3B_1(n \ra \pis)$				&	4.20	&	4.20	&	4.55(37)&	2.18(27)	\\
Triazine			&	$^1A_1''(n \ra \pis)$			&	4.85	&	4.84	&	4.77(13)&	5.12(51)	\\
					&	$^3A_2''(n \ra \pis)$			&	4.40	&	4.40	&	4.45(39)&	4.73(6)		\\
%\hiderowcolors
\hline  % Please only put a hline at the end of the table
\end{tabular}
\begin{tablenotes}
\item $^a$ Excitation energies and error bars \alert{(in eV)} estimated via the novel statistical method based on Gaussian random variables (see Sec.~\ref{sec:error}).
The error bars reported in parenthesis correspond to one standard deviation.
\item $^b$ Excitation energies obtained via a three-point linear fit using the three largest CIPSI variational wave functions, and error bars \alert{(in eV)} estimated via the extrapolation distance, \ie, the difference in excitation energies obtained with the three-point linear extrapolation and the largest CIPSI wave function.
\end{tablenotes}
\end{threeparttable}
\end{table}

%%% FIGURE 3 %%%

%%%%%%%%%%%%%%%%%%%%%%%%%%%%%
\section{The QUEST database}
\label{sec:QUEST}
%%%%%%%%%%%%%%%%%%%%%%%%%%%%%

%=======================
\subsection{Overview}
\begin{figure}
	\centering
	\includegraphics[width=\linewidth]{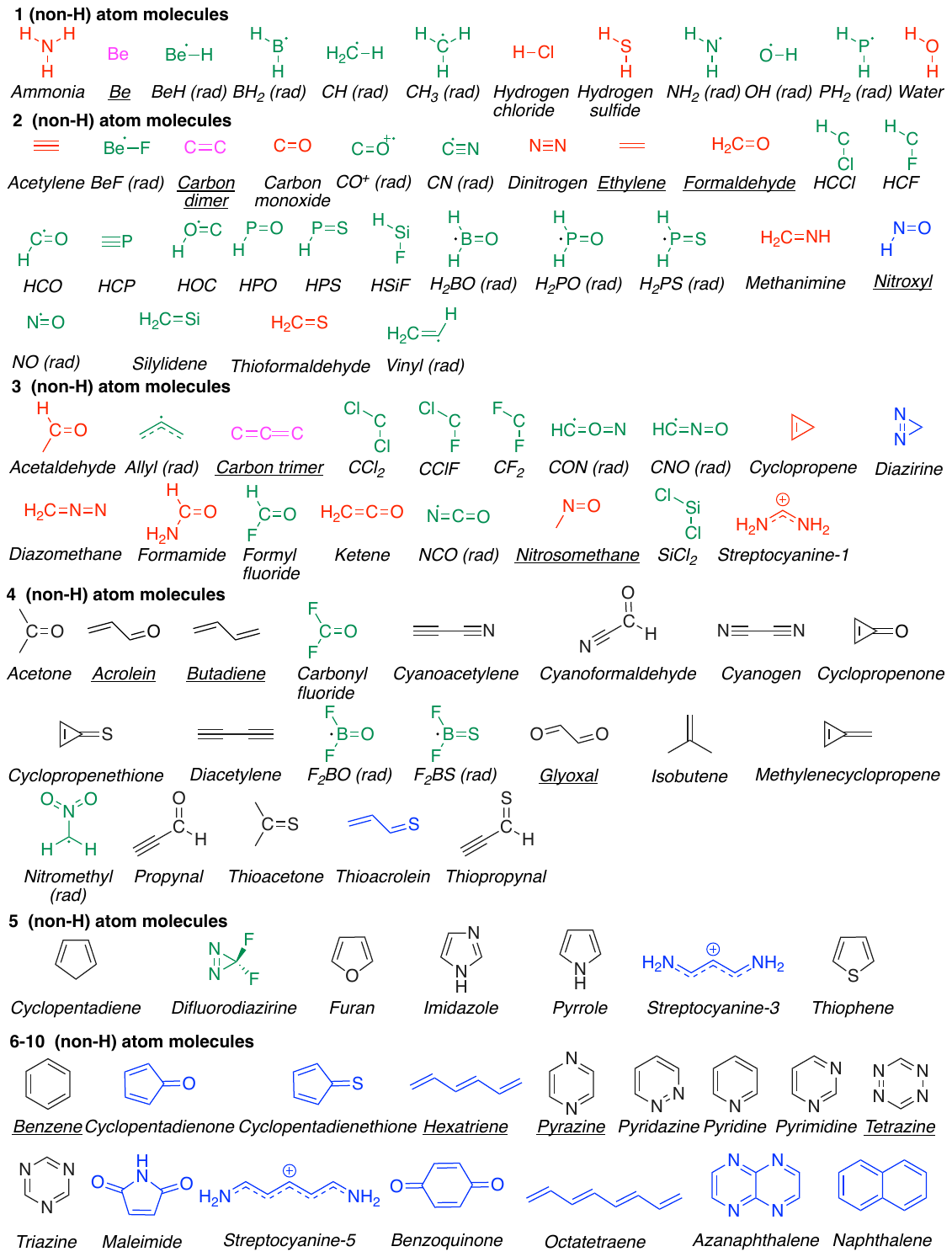}
	\caption{Molecules from each of the five subsets making up the present QUEST dataset of highly-accurate vertical excitation energies:
	QUEST\#1 (red), QUEST\#2 (magenta and/or underlined), QUEST\#3 (black), QUEST\#4 (green), and QUEST\#5 (blue).}
	\label{fig:molecules}
\end{figure}

%=======================
The QUEST database gathers more than 500 highly-accurate excitation energies of various natures (valence, Rydberg, $n \ra \pis$, $\pi \ra \pis$, singlet, doublet, triplet, and double excitations) for molecules ranging
from diatomics to molecules as large as naphthalene (see Fig.~\ref{fig:molecules}). This set is also chemically diverse, with organic and inorganic systems, open- and closed-shell compounds, acyclic and cyclic systems,
pure hydrocarbons and various heteroatomic structures, etc. Each of the five subsets making up the QUEST dataset is detailed below. Throughout the present review, we report several statistical indicators: the mean signed
error (MSE), mean absolute error (MAE), root-mean square error (RMSE), and standard deviation of the errors (SDE), as well as the maximum positive [Max(+)] and maximum negative [Max($-$)] errors.

%%% FIGURE 4 %%%

%=======================
\subsection{QUEST\#1}
%=======================
The QUEST\#1 benchmark set \cite{Loos_2018a} consists of 110 vertical excitation energies (as well as oscillator strengths) from 18 molecules with sizes ranging from one to three non-hydrogen atoms
(water, hydrogen sulfide, ammonia, hydrogen chloride, dinitrogen, carbon monoxide, acetylene, ethylene, formaldehyde, methanimine, thioformaldehyde, acetaldehyde, cyclopropene, diazomethane,
formamide, ketene, nitrosomethane, and the smallest streptocyanine). For this set, we provided two sets of TBEs: i) one obtained within the frozen-core approximation and the aug-cc-pVTZ basis set, and ii)
another one including further corrections for basis set incompleteness and ``all electron'' effects. For the former set, we systematically employed FCI/aug-cc-pVTZ values to define our TBEs, except for a few cases.
For the latter set, both the ``all electron'' correlation and the basis set corrections were systematically obtained at the CC3 level of theory and with the d-aug-cc-pV5Z basis for the nine smallest molecules, and
slightly more compact basis sets for the larger compounds. Our TBE/aug-cc-pVTZ reference excitation energies were employed to benchmark a series of popular excited-state wave function methods partially
or fully accounting for double and triple excitations, namely CIS(D), CC2, CCSD, STEOM-CCSD, CCSDR(3), CCSDT-3, CC3, ADC(2), and ADC(3).  Our main conclusions were that i) ADC(2) and CC2 show
strong similarities in terms of accuracy, ii) STEOM-CCSD is, on average, as accurate as CCSD, the latter overestimating transition energies, iii) CC3 is extremely accurate (with a mean absolute error of only
$\sim 0.03$ eV) and that although slightly less accurate than CC3, CCSDT-3 could be used as a reliable reference for benchmark studies, and iv) ADC(3) was found to be significantly less accurate than CC3
by overcorrecting ADC(2) excitation energies.

%=======================
\subsection{QUEST\#2}
%=======================
The QUEST\#2 benchmark set \cite{Loos_2019} reports reference energies for double excitations. This set gathers 20 vertical transitions from 14 small- and medium-sized molecules (acrolein, benzene, beryllium atom,
butadiene, carbon dimer and trimer, ethylene, formaldehyde, glyoxal, hexatriene, nitrosomethane, nitroxyl, pyrazine, and tetrazine). The TBEs of the QUEST\#2 set are obtained with SCI and/or multiconfigurational
[CASSCF, CASPT2, (X)MS-CASPT2, and NEVPT2] calculations depending on the size of the molecules and the level of theory that we could afford. An important addition to this second study was also the inclusion of
various flavors of multiconfigurational methods (CASSCF, CASPT2, and NEVPT2) in addition to high-order CC methods including, at least, perturbative triples (CC3, CCSDT, CCSDTQ, etc).
Our results demonstrated that the error of CC methods is intimately linked to the amount of double-excitation character in the vertical transition. For ``pure'' double excitations (i.e., for transitions which do not mix with
single excitations), the error in CC3 and CCSDT can easily reach $1$ and $0.5$ eV, respectively, while it goes down to a few tenths of an eV for more common transitions involving a significant amount of single excitations
(such as the well-known $A_g$ transition in butadiene or the $E_{2g}$ excitation in benzene). The quality of the excitation energies obtained with CASPT2 and NEVPT2 was harder to predict as the overall accuracy of
these methods is highly dependent on both the system and the selected active space. Nevertheless, these two methods were found to be more accurate for transitions with a very small percentage of single excitations
(error usually below $0.1$ eV) than for excitations dominated by single excitations where the error is closer to $0.1$--$0.2$ eV.

%=======================
\subsection{QUEST\#3}
%=======================
The QUEST\#3 benchmark set \cite{Loos_2020b} is, by far, our largest set, and consists of highly accurate vertical transition energies and oscillator strengths obtained for 27 molecules encompassing 4, 5, and
6 non-hydrogen atoms (acetone, acrolein, benzene, butadiene, cyanoacetylene, cyanoformaldehyde, cyanogen, cyclopentadiene, cyclopropenone, cyclopropenethione, diacetylene, furan, glyoxal, imidazole, isobutene,
methylenecyclopropene, propynal, pyrazine, pyridazine, pyridine, pyrimidine, pyrrole, tetrazine, thioacetone, thiophene, thiopropynal, and triazine) for a total of 238 vertical transition energies and 90 oscillator strengths
with a reasonably good balance between singlet, triplet, valence, and Rydberg excited states. For these 238 transitions, we have estimated that 224 are chemically accurate for the aug-cc-pVTZ basis and for the
considered geometry. To define the TBEs of the QUEST\#3 set, we employed CC methods up to the highest technically possible order (CC3, CCSDT, and CCSDTQ), and, when affordable SCI calculations with very
large reference spaces (up to hundred million determinants in certain cases), as well as one of the most reliable multiconfigurational methods, NEVPT2, for double excitations. Most of our TBEs are based on CCSDTQ
(4 non-hydrogen atoms) or CCSDT (5 and 6 non-hydrogen atoms) excitation energies. For all the transitions of the QUEST\#3 set, we reported at least CCSDT/aug-cc-pVTZ (sometimes with basis set extrapolation)
and CC3/aug-cc-pVQZ transition energies as well as CC3/aug-cc-pVTZ oscillator strengths for each dipole-allowed transition. Pursuing our previous benchmarking efforts, we confirmed that CC3 almost systematically
delivers transition energies in agreement with higher-level theoretical models ($\pm0.04$ eV) except for transitions presenting a dominant double-excitation character where multiconfigurational methods like NEVPT2 have
logically the edge. This settles down, at least for now, the debate by demonstrating the superiority of CC3 (in terms of accuracy) compared to methods like CCSDT-3 or ADC(3). For the latter model, this was further
demonstrated in a recent study by two of the present authors \cite{Loos_2020d}.

%=======================
\subsection{QUEST\#4}
%=======================
The QUEST\#4 benchmark set \cite{Loos_2020c} consists of two subsets of excitations and oscillator strengths. An ``exotic'' subset of 30 excited states for closed-shell molecules containing F, Cl, P, and Si atoms
(carbonyl fluoride, \ce{CCl2}, \ce{CClF}, \ce{CF2}, difluorodiazirine, formyl fluoride, \ce{HCCl}, \ce{HCF}, \ce{HCP}, \ce{HPO}, \ce{HPS}, \ce{HSiF}, \ce{SiCl2}, and silylidene) and a ``radical'' subset of 51 doublet-doublet
transitions in 24 small radicals (allyl, \ce{BeF}, \ce{BeH}, \ce{BH2}, \ce{CH}, \ce{CH3}, \ce{CN}, \ce{CNO}, \ce{CON}, \ce{CO+}, \ce{F2BO}, \ce{F2BS}, \ce{H2BO}, \ce{HCO}, \ce{HOC}, \ce{H2PO}, \ce{H2PS}, \ce{NCO},
\ce{NH2}, nitromethyl, \ce{NO}, \ce{OH}, \ce{PH2}, and vinyl) characterized by open-shell electronic configurations and an unpaired electron. This represents a total of 81 high-quality TBEs, the vast majority being obtained
at the FCI level with at least the aug-cc-pVTZ basis set. We additionnaly performed high-order CC calculations to ascertain these estimates.  For the exotic set, these TBEs have been used to assess the performances of
15 ``lower-order'' wave function approaches, including several CC and ADC variants. Consistent with our previous works, we found that CC3 is very accurate, whereas the trends for the other methods are similar to that
obtained on more standard CNOSH organic compounds. In contrast, for the radical set, even the refined ROCC3 method yields a comparatively large MAE of $0.05$ eV. Likewise, the excitation energies obtained with CCSD
are much less satisfying for open-shell derivatives (MAE of $0.20$ eV with UCCSD and $0.15$ eV with ROCCSD) than for closed-shell systems of similar size (MAE of $0.07$ eV).

%=======================
\subsection{QUEST\#5}
%=======================

The QUEST\#5 subset is composed of additional accurate excitation energies that we have produced for the present article. This new set gathers 13 new systems composed by small molecules as well as larger molecules
(see blue molecules in Fig.~\ref{fig:molecules}): aza-naphthalene, benzoquinone, cyclopentadienone, cyclopentadienethione, diazirine, hexatriene, maleimide, naphthalene, nitroxyl, octatetraene, streptocyanine-C3, streptocyanine-C5,
and thioacrolein. For these new transitions, we report again quality vertical transition energies, the vast majority being of CCSDT quality, and we consider that, out of these 80 new transitions, 55 of them can be labeled
as ``safe'', \ie, considered as chemically accurate or within 0.05 eV of the FCI limit for the given geometry and basis set. We refer the interested reader to the {\SupInf} for a detailed discussion of each molecule for which comparisons
are made with literature data.

%%%%%%%%%%%%%%%%%%%%%%%%%%%%%
\section{Theoretical best estimates}
\label{sec:TBE}
%%%%%%%%%%%%%%%%%%%%%%%%%%%%%
We discuss in this section the generation of the TBEs obtained with the aug-cc-pVTZ basis.
For the closed-shell compounds, the exhaustive list of TBEs can be found in Table \ref{tab:TBE} alongside various specifications: the molecule's name, the excitation, its nature (valence, Rydberg, or charge transfer), its oscillator strength (when symmetry- and spin-allowed),
and its percentage of single excitations $\%T_1$ (computed at the LR-CC3 level). All these quantities are computed with the same aug-cc-pVTZ basis.
Importantly, we also report the composite approach considered to compute the TBEs (see column ``Method'').
Following an ONIOM-like strategy \cite{Svensson_1996a,Svensson_1996b}, the TBEs are computed as ``A/SB + [B/TB - B/SB]'', where A/SB is the excitation energy computed with a method A in a smaller basis (SB), and B/SB and B/TB are excitation energies computed with a method B in the small basis and target basis TB, respectively.
Table \ref{tab:rad} reports the TBEs for the open-shell molecules belonging to the QUEST\#4 subset.

Talking about numbers, the QUEST database is composed of 551 excitation energies, including 302 singlet, 197 triplet, 51 doublet, 412 valence, and 176 Rydberg excited states.
Amongst the valence transitions in closed-shell compounds, 135 transitions correspond to $n \ra \pis$ excitations, 200 to $\pi \ra \pis$ excitations, and 23 are doubly-excited states. In terms of molecular sizes, 146 excitations are obtained
in molecules having in-between 1 and 3 non-hydrogen atoms, 97 excitations from 4 non-hydrogen atom compounds, 177 from molecules composed by 5 and 6 non-hydrogen atoms, and, finally, 68 excitations are obtained from systems with 7 to 10 non-hydrogen atoms.
In addition, QUEST is composed by 24 open-shell molecules with a single unpaired electron.
Amongst these excited states, 485 of them are considered as ``safe'', \ie, chemically-accurate for the considered basis set and geometry.
Besides this energetic criterion, we consider as ``safe'' transitions that are either: i) computed with FCI or CCSDTQ, or ii) in which the difference between CC3 and CCSDT excitation energies is small (\ie, around $0.03$--$0.04$ eV) with a large $\%T_1$ value.

\begin{center}
\scriptsize
\begin{longtable}{clccccclc}
\caption{Theoretical best estimates TBEs (in eV), oscillator strengths $f$, percentage of single excitations $\%T_1$ involved in the transition (computed at the CC3 level) for the full set of closed-shell compounds of the QUEST database.
``Method'' provides the protocol employed to compute the TBEs.
The nature of the excitation is also provided: V, R, and CT stands for valence, Rydberg, and charge transfer, respectively.
[F] indicates a fluorescence transition, \ie, a vertical transition energy computed from an excited-state geometry.
AVXZ stands for aug-cc-pVXZ.
\label{tab:TBE}}
\\
\hline
\#	&\thead{Molecule}	&	\thead{Excitation}	&	\thead{Nature}		&	\thead{$\%T_1$}	&	\thead{$f$}	&	\thead{TBE}	&	\thead{Method} &	\thead{Safe?}\\
\hline
\endfirsthead
\multicolumn{9}{c}{\tablename\ \thetable\ -- \textit{Continued from previous page}} \\
\hline
\thead{\#}	&\thead{Molecule}	&	\thead{Excitation}	&	\thead{Nature}		&	\thead{$\%T_1$}	&	\thead{$f$}	&	\thead{TBE}	&	\thead{Method} &	\thead{Safe?}\\
\hline
\endhead
\hline \multicolumn{9}{r}{\textit{Continued on next page}} \\
\endfoot
\hline
\endlastfoot
1	&	Acetaldehyde	&	$^1A''  (n \ra \pi^*)$	&	V	&	91	&	0.000	&	4.31	&	FCI/AVTZ	&	Y	\\
2	&		&	$^3A''  (n \ra \pi^*)$	&	V	&	97	&		&	3.97	&	FCI/AVDZ + [CCSDT/AVTZ - CCSDT/AVDZ]	&	Y	\\
3	&	Acetone	&	$^1A_2   (n \ra \pi^*)$	&	V	&	91	&		&	4.47	&	CCSDTQ/6-31+G(d) + [CCSDT/AVTZ - CCSDT/6-31+G(d)]	&	Y	\\
4	&		&	$^1B_2   (n \ra 3s)$	&	R	&	90	&	0.000	&	6.46	&	CCSDTQ/6-31+G(d) + [CCSDT/AVTZ - CCSDT/6-31+G(d)]	&	Y	\\
5	&		&	$^1A_2   (n \ra 3p)$	&	R	&	90	&		&	7.47	&	CCSDTQ/6-31+G(d) + [CCSDT/AVTZ - CCSDT/6-31+G(d)]	&	Y	\\
6	&		&	$^1A_1   (n \ra 3p)$	&	R	&	90	&	0.004	&	7.51	&	CCSDTQ/6-31+G(d) + [CCSDT/AVTZ - CCSDT/6-31+G(d)]	&	Y	\\
7	&		&	$^1B_2   (n \ra 3p)$	&	R	&	91	&	0.029	&	7.62	&	CCSDTQ/6-31+G(d) + [CCSDT/AVTZ - CCSDT/6-31+G(d)]	&	Y	\\
8	&		&	$^3A_2   (n \ra \pi^*)$	&	V	&	97	&		&	4.13	&	CCSDT/AVDZ + [CC3/AVTZ - CC3/AVDZ]	&	Y	\\
\end{longtable}
\end{center}

%%% TABLE III %%%
\begin{table}[htp]
\centering
\scriptsize
\caption{Theoretical best estimates TBEs (in eV) for the doublet-doublet transitions of the open-shell molecules belonging to QUEST\#4.
These TBEs are obtained with the aug-cc-pVTZ basis set, and ``Method'' indicates the protocol employed to compute them.}
\label{tab:rad}
\begin{threeparttable}
\begin{tabular}{cllcl}
\headrow
\thead{\#} 	& \thead{Molecule} 	& \thead{Transition} 	& \thead{TBE/aug-cc-pVTZ}	&	\thead{Method}	\\
1	&	Allyl			&$^2B_1$		&3.39	&	FCI/6-31+G(d) + [CCSDT/aug-cc-pVTZ - CCSDT/6-31+G(d)]	\\
2	&					&$^2A_1$		&4.99	&	FCI/6-31+G(d) + [CCSDT/aug-cc-pVTZ - CCSDT/6-31+G(d)]	\\
3	&	\ce{BeF}		&$^2\Pi$		&4.14	&	FCI/aug-cc-pVTZ	\\
4	&					&$^2\Sigma^+$	&6.21   &   FCI/aug-cc-pVTZ	\\
5	&	\ce{BeH}		&$^2\Pi$		&2.49	&   FCI/aug-cc-pVTZ	\\
6	&					&$^2\Pi$		&6.46	&   FCI/aug-cc-pVTZ	\\
7	&	\ce{BH2}		&$^2B_1$		&1.18	&   FCI/aug-cc-pVTZ	\\
\hline
\end{tabular}
\end{threeparttable}
\end{table}
%%% %%% %%% %%%

%%%%%%%%%%%%%%%%%%%%%%%%%%%%%
\section{Benchmarks}
\label{sec:bench}
\begin{figure}
	\centering
	\includegraphics[width=0.9\textwidth]{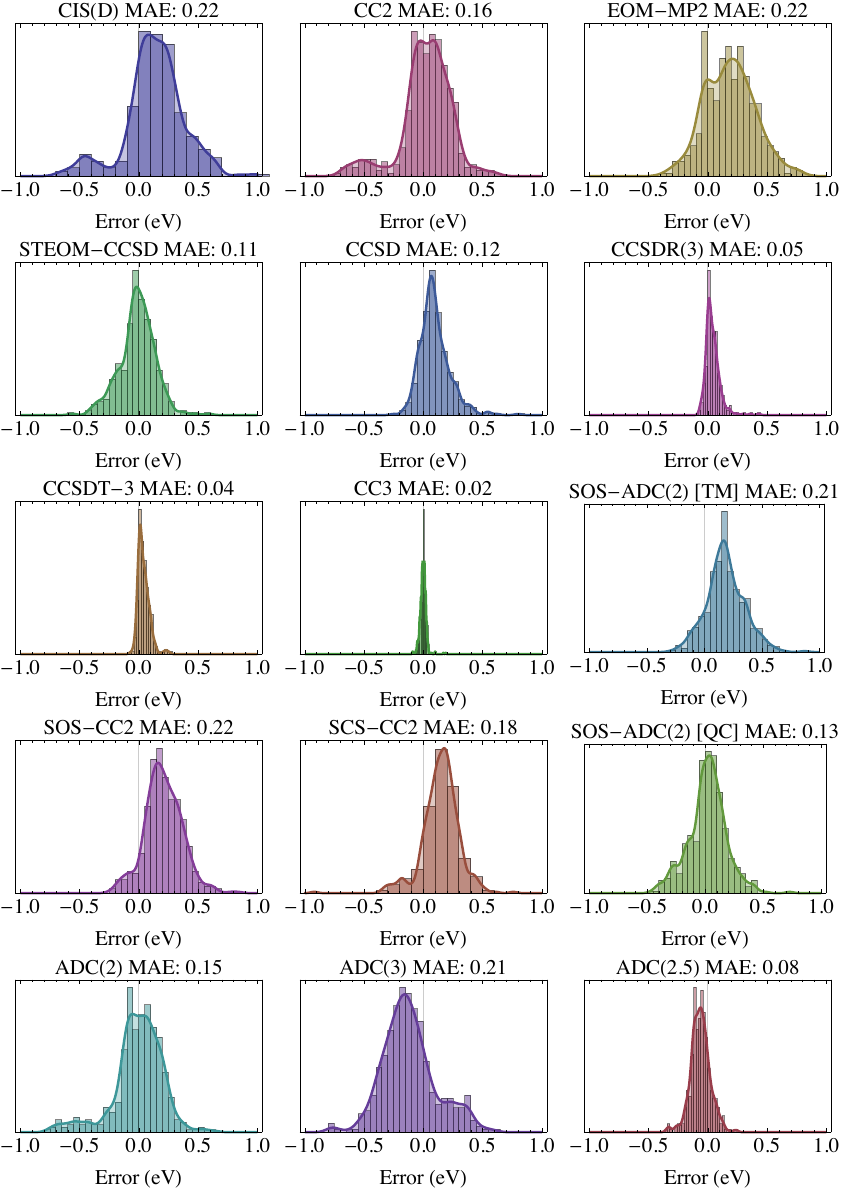}
	\caption{Distribution of the error (in eV) in excitation energies (with respect to the TBE/aug-cc-pVTZ values) for various methods for the entire QUEST database considering only closed-shell compounds.
	Only the ``safe'' TBEs are considered (see Table \ref{tab:TBE}).
	See Table \ref{tab:stat} for the values of the corresponding statistical quantities.
	QC and TM indicate that Q-CHEM and TURBOMOLE scaling factors are considered, respectively. The SOS-CC2 and SCS-CC2 approaches are obtained with the latter code.
	\label{fig:QUEST_stat}}
\end{figure}

%%%%%%%%%%%%%%%%%%%%%%%%%%%%%
In this section, we report a comprehensive benchmark of various lower-order methods on the entire set of closed-shell compounds belonging to the QUEST database.
Statistical quantities are reported in Table \ref{tab:stat} (the entire set of data can be found in the {\SupInf}).
Additionally, we also provide a specific analysis for each type of excited states.
Hence, the statistical values are reported for various types of excited states and molecular sizes for the MSE and MAE.
The distribution of the errors in vertical excitation energies (with respect to the TBE/aug-cc-pVTZ reference values) are represented in Fig.~\ref{fig:QUEST_stat} for all the ``safe'' excitations having a dominant single excitation character (\ie, the double excitations are discarded).
Similar graphs are reported in the {\SupInf} for specific sets of transitions and molecules.
\alert{For the vast majority of the cases, the comparison between methods presented here is performed for Franck-Condon geometries only.
Therefore, it is important to stress that the present method ranking might change significantly when moving away from the Franck-Condon region as most excited-state methods do not provide a uniform description of potential energy surfaces.}

%%% TABLE IV %%%
\begin{sidewaystable}
\scriptsize
\centering
\caption{Mean signed error (MSE), mean absolute error (MAE), root-mean-square error (RMSE), standard deviation of the errors (SDE), as well as the maximum positive error [Max(+)] and negative error [Max($-$)] with respect to the TBE/aug-cc-pVTZ for the entire QUEST database.
Only the ``safe'' TBEs are considered (see Table \ref{tab:TBE}).
For the MSE and MAE, the statistical values are reported for various types of excited states and molecular sizes.
All quantities are given in eV.
``Count'' refers to the number of transitions considered for each method.}
\label{tab:stat}
\begin{threeparttable}
\begin{tabular}{llccccccccccccccc}
\headrow
			&				&	\thead{CIS(D)} &	\thead{CC2}	& \thead{EOM-MP2} &	\thead{STEOM-CCSD}	& \thead{CCSD}	& \thead{CCSDR(3)}	& \thead{CCCSDT-3}	& \thead{CC3}
							&	\thead{SOS-ADC(2)$^a$}	& \thead{SOS-CC2$^a$}	& \thead{SCS-CC2$^a$}	& \thead{SOS-ADC(2)$^b$}	& \thead{ADC(2)}	& \thead{ADC(3)}	& \thead{ADC(2.5)} \\
Count		&				& 429	& 431	& 427	& 360	& 431	& 259	& 251	& 431	& 430	& 430	& 430	& 430	& 426	& 423	& 423	\\
Max(+)		&				& 1.06	& 0.63	& 0.80	& 0.59	& 0.80	& 0.43	& 0.26	& 0.19	& 0.87	& 0.84	& 0.76	& 0.73	& 0.64	& 0.60	& 0.24	\\
Max($-$)	&				& -0.69	& -0.71	& -0.38	& -0.56	& -0.25	& -0.07	& -0.07	& -0.09	& -0.29	& -0.24	& -0.92	& -0.46	& -0.76	& -0.79	& -0.34	\\
MSE			&				& 0.13	& 0.02	& 0.18	& -0.01	& 0.10	& 0.04	& 0.04	& 0.00	& 0.18	& 0.21	& 0.15	& 0.02	& -0.01	& -0.12	& -0.06	\\
			& singlet		& 0.10	& -0.02	& 0.22	& 0.03	& 0.14	& 0.04	& 0.04	& 0.00	& 0.18	& 0.20	& 0.13	& 0.00	& -0.04	& -0.08	& -0.06	\\
			& triplet		& 0.19	& 0.08	& 0.14	& -0.07	& 0.03	& 		& 		& 0.00	& 0.19	& 0.22	& 0.17	& 0.04	& 0.04	& -0.18	& -0.07	\\
			& valence		& 0.20	& 0.10	& 0.20	& -0.06	& 0.10	& 0.06	& 0.05	& 0.00	& 0.19	& 0.24	& 0.20	& 0.02	& 0.04	& -0.16	& -0.06	\\
			& Rydberg		& -0.04	& -0.17	& 0.15	& 0.09	& 0.08	& 0.01	& 0.03	& -0.01	& 0.16	& 0.12	& 0.01	& 0.02	& -0.13	& -0.02	& -0.07	\\
			& $n \ra \pis$	& 0.16	& 0.02	& 0.24	& -0.03	& 0.17	& 0.07	& 0.07	& 0.00	& 0.26	& 0.32	& 0.22	& 0.05	& -0.05	& -0.01	& -0.03	\\
			& $\pi \ra \pis$& 0.25	& 0.17	& 0.20	& -0.07	& 0.06	& 0.05	& 0.04	& 0.00	& 0.15	& 0.19	& 0.19	& 0.00	& 0.12	& -0.27	& -0.07	\\
			& 1--3 non-H	& 0.10	& 0.03	& 0.03	& -0.02	& 0.04	& 0.01	& 0.01	& 0.00	& 0.13	& 0.16	& 0.11	& -0.01	& -0.01	& -0.17	& -0.09	\\
			& 4 non-H		& 0.13	& 0.04	& 0.12	& 0.00	& 0.09	& 0.03	& 0.04	& 0.00	& 0.19	& 0.26	& 0.19	& 0.03	& -0.04	& -0.10	& -0.07	\\
			& 5--6 non-H	& 0.17	& 0.02	& 0.30	& -0.01	& 0.11	& 0.05	& 0.05	& 0.00	& 0.21	& 0.20	& 0.14	& 0.03	& 0.03	& -0.10	& -0.04	\\
			& 7--10 non-H	& 0.15	& -0.03	& 0.42	& -0.05	& 0.22	& 0.10	& 0.08	& -0.01	& 0.26	& 0.29	& 0.19	& 0.05	& -0.06	& -0.02	& -0.04	\\
SDE			&				& 0.24	& 0.20	& 0.21	& 0.13	& 0.12	& 0.05	& 0.04	& 0.02	& 0.17	& 0.16	& 0.16	& 0.15	& 0.20	& 0.22	& 0.08	\\
RMSE		&				& 0.29	& 0.22	& 0.28	& 0.15	& 0.16	& 0.07	& 0.06	& 0.03	& 0.25	& 0.26	& 0.22	& 0.17	& 0.21	& 0.26	& 0.10	\\
MAE			&				& 0.22	& 0.16	& 0.22	& 0.11	& 0.12	& 0.05	& 0.04	& 0.02	& 0.20	& 0.22	& 0.18	& 0.13	& 0.15	& 0.21	& 0.08	\\
			& singlet		& 0.22	& 0.16	& 0.25	& 0.10	& 0.14	& 0.05	& 0.04	& 0.02	& 0.21	& 0.22	& 0.17	& 0.14	& 0.16	& 0.20	& 0.09	\\
			& triplet		& 0.23	& 0.15	& 0.18	& 0.12	& 0.08	& 		& 		& 0.01	& 0.20	& 0.23	& 0.19	& 0.11	& 0.15	& 0.22	& 0.08	\\
			& valence		& 0.22	& 0.14	& 0.24	& 0.12	& 0.13	& 0.06	& 0.05	& 0.02	& 0.21	& 0.25	& 0.20	& 0.12	& 0.13	& 0.22	& 0.08	\\
			& Rydberg		& 0.22	& 0.21	& 0.19	& 0.10	& 0.08	& 0.03	& 0.03	& 0.02	& 0.20	& 0.15	& 0.13	& 0.14	& 0.21	& 0.18	& 0.09	\\
			& $n \ra \pis$	& 0.18	& 0.08	& 0.28	& 0.08	& 0.17	& 0.07	& 0.07	& 0.01	& 0.26	& 0.32	& 0.22	& 0.11	& 0.10	& 0.14	& 0.07	\\
			& $\pi \ra \pis$& 0.27	& 0.19	& 0.21	& 0.14	& 0.11	& 0.06	& 0.04	& 0.02	& 0.18	& 0.21	& 0.20	& 0.12	& 0.16	& 0.28	& 0.09	\\
			& 1--3 non-H	& 0.23	& 0.19	& 0.13	& 0.10	& 0.07	& 0.03	& 0.03	& 0.02	& 0.18	& 0.20	& 0.19	& 0.14	& 0.19	& 0.24	& 0.10	\\
			& 4 non-H		& 0.22	& 0.19	& 0.15	& 0.11	& 0.11	& 0.03	& 0.04	& 0.02	& 0.19	& 0.26	& 0.22	& 0.13	& 0.18	& 0.23	& 0.08	\\
			& 5--6 non-H	& 0.21	& 0.12	& 0.30	& 0.12	& 0.13	& 0.06	& 0.05	& 0.01	& 0.22	& 0.21	& 0.15	& 0.11	& 0.11	& 0.19	& 0.07	\\
			& 7--10 non-H	& 0.24	& 0.11	& 0.42	& 0.12	& 0.23	& 0.10	& 0.08	& 0.02	& 0.27	& 0.29	& 0.19	& 0.12	& 0.14	& 0.16	& 0.07	\\
\hline
\end{tabular}
\begin{tablenotes}
\item $^a$ Excitation energies computed with TURBOMOLE.
\item $^b$ Excitation energies computed with Q-CHEM.
\end{tablenotes}
\end{threeparttable}
\end{sidewaystable}

%%% FIGURE 5 %%%

The most striking feature from the statistical indicators gathered in Table \ref{tab:stat} is the overall accuracy of CC3 with MAEs and MSEs systematically below the chemical accuracy threshold (errors $<$ 0.043 eV or 1 kcal/mol), irrespective of the nature of the transition and the size of the molecule.
CCSDR(3) are CCCSDT-3 can also be regarded as excellent performers with overall MAEs below $0.05$ eV, though one would notice a slight degradation of their performances for the $n  \ra \pis$ excitations and the largest molecules of the database.
The other third-order method, ADC(3), which enjoys a lower computational cost, is significantly less accurate and does not really improve upon its second-order analog, even for the largest systems considered here, an observation in line with a previous analysis by some of the authors \cite{Loos_2020d}.
Nonetheless, ADC(3)'s accuracy improves in larger compounds, with a MAE of 0.24 eV (0.16 eV) for the subsets of the most compact (extended) compounds considered herein. The ADC(2.5) composite method introduced in Ref.~\cite{Loos_2020d}, which corresponds to grossly average the ADC(2) and ADC(3)
values, yields an appreciable accuracy improvement, as shown in Fig.~\ref{fig:QUEST_stat}.  Indeed, we note that the MAE of 0.07 eV obtained for ``large'' compounds is comparable to the one obtained with CCSDR(3) and CCSDT-3 for these molecules.  All these third-order methods
are rather equally efficient for valence and Rydberg transitions.

Concerning the second-order methods (which have the indisputable advantage to be applicable to larger molecules than the ones considered here), we have the following ranking in terms of MAEs: EOM-MP2 $\approx$ CIS(D) $<$ CC2 $\approx$ ADC(2) $<$ CCSD $\approx$ STEOM-CCSD, which fits our previous conclusions on the specific subsets \cite{Loos_2018a,Loos_2019,Loos_2020b,Loos_2020c,Loos_2020d}.
A very similar ranking is obtained when one looks at the MSEs.
It is noteworthy that the performances of EOM-MP2 and CCSD are getting notably worse when the system size increases, while CIS(D) and STEOM-CCSD have a very stable behavior with respect to system size.
Indeed, the EOM-MP2 MAE attains 0.42 eV for molecules containing between 7 and 10 non-hydrogen atoms, whereas the CCSD tendency to overshoot the transition energies yield a MSE of 0.22 eV for the same set (a rather large error).

For CCSD, this conclusion fits benchmark studies published by other groups \cite{Schreiber_2008,Caricato_2010,Watson_2013,Kannar_2014,Kannar_2017,Dutta_2018}.
For example, K\'ann\'ar and Szalay obtained a MAE of 0.18 eV on Thiel's set for the states exhibiting a dominant single excitation character.
The CCSD degradation with system size might partially explain the similar (though less pronounced) trend obtained for CCSDR(3).
Regarding the apparently better performances of STEOM-CCSD as compared to CCSD, we recall that several challenging states have been naturally removed from the STEOM-CCSD statistics because the active character percentage was lower than $98\%$ (see above).

In contrast to EOM-MP2 and CCSD, the overall accuracy of CC2 and ADC(2) does significantly improve for larger molecules, the performances of the two methods being, as expected, similar \cite{Harbach_2014}.
Let us note that these two methods show similar accuracies for singlet and triplet transitions, but are significantly less accurate for Rydberg transitions, as already pointed out previously \cite{Kannar_2017}.
Therefore, both CC2 and ADC(2) offer an appealing cost-to-accuracy ratio for large compounds, which explains their popularity in realistic chemical scenarios \cite{Hattig_2005c,Goerigk_2010a,Send_2011a,Winter_2013,Jacquemin_2015b,Oruganti_2016}.
For the scaled methods [SOS-ADC(2), SOS-CC2, and SCS-CC2], the TURBOMOLE scaling factors do not seem to improve things upon the unscaled versions, while the Q-CHEM scaling factors for ADC(2) provide a small, yet significant improvement for this set of molecules.
Of course, one of the remaining open questions regarding all these methods is their accuracy for even larger systems.

%%%%%%%%%%%%%%%%%%%%%%%%%%%%%
\section{The QUESTDB website}
\label{sec:website}
%%%%%%%%%%%%%%%%%%%%%%%%%%%%%

Quite a large number of calculations were required for each of the
QUEST articles \cite{Loos_2018a,Loos_2019,Loos_2020b,Loos_2020c,Loos_2020d}.
Up to now, all the curated data was shared as
supplementary information presented as a file in portable document
format (pdf). This way of sharing data does not require too much
effort for the authors, but it is obviously not optimal from the
user's point of view.
We have now addressed this problem by creating a database which
contains all the vertical and fluorescence transition energies as well
the corresponding molecular geometries. This data can be manipulated via
a web application which allows to plot the statistical indicators (generated with the \texttt{Plotly} library)
computed on selected subsets of molecules, methods and basis sets.
The application also gives the possibility to the user to import
external data files, in order to compare the performance of methods
that are not in our database.
Both the web application and the data are hosted in a single GitHub
repository (\url{https://github.com/LCPQ/QUESTDB_website})
and available at the following address: \url{https://lcpq.github.io/QUESTDB_website}.
In this way, extending the database is as simple as adding new data files to the
repository, together with the corresponding bibliographic references,
and we strongly encourage users to contribute to enlarge this database
via GitHub pull requests.

%%%%%%%%%%%%%%%%%%%%%%%%%%%%%
\section{Concluding remarks}
\label{sec:ccl}
%%%%%%%%%%%%%%%%%%%%%%%%%%%%%
In the present review article, we have presented and extended the QUEST database of highly-accurate excitation energies for molecular systems \cite{Loos_2020a,Loos_2018a,Loos_2019,Loos_2020b,Loos_2020c} that we started building
in 2018 and that is now composed by more than 500 vertical excitations, many of which can be reasonably considered as within 1 kcal/mol (or less) of the FCI limit for the considered CC3/aug-cc-pVTZ geometry and basis set (\emph{aug}-cc-pVTZ).
In particular, we have detailed the specificities of our protocol by providing computational details regarding geometries, basis sets, as well as reference and benchmarked computational methods. The content of our five QUEST subsets has
been presented in detail, and for each of them, we have provided the number of reference excitation energies, the nature and size of the molecules, the list of benchmarked methods, as well as other useful specificities.
Importantly, we have proposed a new statistical method that produces much safer estimates of the extrapolation error in SCI calculations. This new method based on Gaussian random variables has been tested by computing additional FCI values for five- and six-membered rings.
After having discussed the generation of our TBEs, we have reported a comprehensive benchmark for a significant number of methods on the entire QUEST set with, in addition, a specific analysis for each type of excited states.
Finally, the main features of the website specifically designed to gather the entire data generated during these past few years have been presented and discussed.

Paraphrasing Thiel's conclusions \cite{Schreiber_2008}, we hope that not only the QUEST database will be used for further benchmarking and testing, but that other research groups will also improve it, providing not only corrections
(inevitable in such a large data set), but more importantly extensions with both improved estimates for some compounds and states, or new molecules.
In this framework, we provide in the {\SupInf} a file with all our benchmark data.

Regarding future improvements and extensions, we would like to mention that although our present goal is to produce chemically accurate vertical excitation energies, we are currently devoting great efforts to obtain highly-accurate excited-state properties \cite{Hodecker_2019,Eriksen_2020b} such as dipoles and oscillator strengths for molecules of small and medium sizes \cite{Chrayteh_2021,Sarkar_2021}, so as to complete previous efforts aiming at determining accurate excited-state geometries \cite{Budzak_2017,Jacquemin_2018}.
\alert{In this context, methods for which one has access to analytic nuclear gradients [\eg, ADC(2), CC2, and EOM-CCSD] and frequencies [\eg, TD-DFT] have an indisputable edge.}
Reference ground-state properties (such as correlation energies and atomization energies) are also being currently produced \cite{Scemama_2020,Loos_2020e}.
\alert{Additional reference energies for charge-transfer excited states \cite{Schwabe_2017,Kozma_2020} and transition metal compounds \cite{Zhao_2006,Williams_2020} would be a valuable addition to the present database.}
Besides this, because computing 500 (or so) excitation energies can be a costly exercise even with cheap computational methods, we are planning on developing a ``diet set'' (\ie, a much smaller set of excitation energies which can reproduce key results of the full QUEST database, including ranking of approximations) following the philosophy of the ``diet GMTKN55'' set proposed recently by Gould \cite{Gould_2018b}.
We hope to report on this in the near future.

%%%%%%%%%%%%%%%%%%%%%%%%%%%%
%\section*{acknowledgements}
%%%%%%%%%%%%%%%%%%%%%%%%%%%%
%AS, MC, and PFL thank the European Research Council (ERC) under the European Union's Horizon 2020 research and innovation programme (Grant agreement No.~863481) for financial support.
%Support from the \textit{``Centre National de la Recherche Scientifique''} is acknowledged.

%%%%%%%%%%%%%%%%%%%%%%%%%%%%
\section*{research resources}
%%%%%%%%%%%%%%%%%%%%%%%%%%%%
This work was performed using HPC resources from GENCI-TGCC (Grand Challenge 2019-gch0418) and from CALMIP (Toulouse) under allocation 2020-18005.
DJ acknowledges the \textit{R\'egion des Pays de la Loire} for financial support and the CCIPL computational center for ultra-generous allocation of computational time.

%%%%%%%%%%%%%%%%%%%%%%%%%%%%%%%%
\section*{conflict of interest}
%%%%%%%%%%%%%%%%%%%%%%%%%%%%%%%%
The authors have declared no conflicts of interest for this article.

%%%%%%%%%%%%%%%%%%%%%%%%%%%%%%%%
\section*{supporting information}
%%%%%%%%%%%%%%%%%%%%%%%%%%%%%%%%
Cartesian coordinates of each molecule (in bohr), Python code associated with the algorithm employed to compute the extrapolated FCI excitation energies and their associated error bars (as well as additional examples for smaller systems), a detailed discussion of each molecule of the QUEST\#5 subset including comparisons with literature data, Excel spreadsheet gathering all benchmark data and additional statistical analyses for various molecular and excitation subsets.

%%%%%%%%%%%%%%%%%%%%%%%%%%%%%%%%
%\bibliography{QUESTDB}

\begin{thebibliography}{233}
\providecommand{\natexlab}[1]{#1}
\providecommand{\url}[1]{\texttt{#1}}
\providecommand{\urlprefix}{}

\bibitem[{Szabo and Ostlund(1989)A. Szabo and N. S. Ostlund}]{SzaboBook}
Szabo A, Ostlund NS.
\newblock Modern quantum chemistry.
\newblock New York: McGraw-Hill; 1989.

\bibitem[{Jensen(2017)F. Jensen}]{JensenBook}
Jensen F.
\newblock Introduction to Computational Chemistry.
\newblock 3rd ed. New York: Wiley; 2017.

\bibitem[{Cramer(2004)C. J. Cramer}]{CramerBook}
Cramer CJ.
\newblock Essentials of Computational Chemistry: Theories and Models.
\newblock Wiley; 2004.

\bibitem[{Helgaker et~al.(2013)T. Helgaker and P. J{\o}rgensen and J.
  Olsen}]{HelgakerBook}
Helgaker T, J{\o}rgensen P, Olsen J.
\newblock Molecular Electronic-Structure Theory.
\newblock John Wiley \& Sons, Inc.; 2013.

\bibitem[{Roos et~al.(1996)Roos, B. O. and Andersson, K. and Fulscher, M. P.
  and Malmqvist, P.-A. and {Serrano-Andr\'es}, L.}]{Roos_1996}
Roos BO, Andersson K, Fulscher MP, Malmqvist PA, {Serrano-Andr\'es} L.
\newblock In: Prigogine I, Rice SA, editors. Multiconfigurational Perturbation
  Theory: Applications In Electronic Spectroscopy, vol. XCIII of Adv. Chem.
  Phys. Wiley, New York; 1996. p. 219--331.

\bibitem[{Piecuch et~al.(2002)Piotr Piecuch and Karol Kowalski and Ian S. O.
  Pimienta and Michael J. Mcguire}]{Piecuch_2002}
Piecuch P, Kowalski K, Pimienta ISO, Mcguire MJ.
\newblock Recent advances in electronic structure theory: Method of moments of
  coupled-cluster equations and renormalized coupled-cluster approaches.
\newblock International Reviews in Physical Chemistry 2002;21:527--655.

\bibitem[{Dreuw and Head-Gordon(2005)Dreuw, Andreas and Head-Gordon,
  Martin}]{Dreuw_2005}
Dreuw A, Head-Gordon M.
\newblock Single-{{Reference}} Ab {{Initio Methods}} for the {{Calculation}} of
  {{Excited States}} of {{Large Molecules}}.
\newblock Chem Rev 2005;105:4009--4037.

\bibitem[{Krylov(2006)Krylov, Anna I.}]{Krylov_2006}
Krylov AI.
\newblock Spin-Flip Equation-of-Motion Coupled-Cluster Electronic Structure
  Method for a Description of Excited States, Bond Breaking, Diradicals, and
  Triradicals.
\newblock Accounts of Chemical Research 2006;39:83--91.
\newblock PMID: 16489727.

\bibitem[{Sneskov and Christiansen(2012)Sneskov, Kristian and Christiansen,
  Ove}]{Sneskov_2012}
Sneskov K, Christiansen O.
\newblock Excited State Coupled Cluster Methods.
\newblock WIREs Comput Mol Sci 2012;2:566--584.

\bibitem[{Gonz{\'a}lez et~al.(2012)Gonz{\'a}lez, Leticia and Escudero, D. and
  Serrano-Andr\`es, L.}]{Gonzales_2012}
Gonz{\'a}lez L, Escudero D, Serrano-Andr\`es L.
\newblock Progress and Challenges in the Calculation of Electronic Excited
  States.
\newblock ChemPhysChem 2012;13:28--51.

\bibitem[{Laurent and Jacquemin(2013)Laurent, Ad{\`e}le D. and Jacquemin,
  Denis}]{Laurent_2013}
Laurent AD, Jacquemin D.
\newblock TD-DFT Benchmarks: A Review.
\newblock Int J Quantum Chem 2013;113:2019--2039.

\bibitem[{Adamo and Jacquemin(2013)Adamo, C. and Jacquemin, D.}]{Adamo_2013}
Adamo C, Jacquemin D.
\newblock The calculations of Excited-State Properties with Time-Dependent
  Density Functional Theory.
\newblock Chem Soc Rev 2013;42:845--856.

\bibitem[{Ghosh et~al.(2018)Ghosh, Soumen and Verma, Pragya and Cramer,
  Christopher J. and Gagliardi, Laura and Truhlar, Donald G.}]{Ghosh_2018}
Ghosh S, Verma P, Cramer CJ, Gagliardi L, Truhlar DG.
\newblock Combining Wave Function Methods with Density Functional Theory for
  Excited States.
\newblock Chem Rev 2018;118:7249--7292.

\bibitem[{Blase et~al.(2020)X. Blase and I. Duchemin and D. Jacquemin and P. F.
  Loos}]{Blase_2020}
Blase X, Duchemin I, Jacquemin D, Loos PF.
\newblock The Bethe-Salpeter Formalism: From Physics to Chemistry.
\newblock J Phys Chem Lett 2020;11:7371.

\bibitem[{Loos et~al.(2020)P. F. Loos and A. Scemama and D.
  Jacquemin}]{Loos_2020a}
Loos PF, Scemama A, Jacquemin D.
\newblock The Quest for Highly-Accurate Excitation Energies: a Computational
  Perspective.
\newblock J Phys Chem Lett 2020;11:2374--2383.

\bibitem[{Bernardi et~al.(1996)Bernardi, Fernando and Olivucci, Massimo and
  Robb, Michael A.}]{Bernardi_1996}
Bernardi F, Olivucci M, Robb MA.
\newblock Potential Energy Surface Crossings in Organic Photochemistry.
\newblock Chem Soc Rev 1996;25:321.

\bibitem[{Olivucci(2010)Olivucci, Massimo}]{Olivucci_2010}
Olivucci M.
\newblock Computational Photochemistry.
\newblock Amsterdam; Boston (Mass.); Paris: {Elsevier Science}; 2010.
\newblock OCLC: 800555856.

\bibitem[{Robb et~al.(2007)Robb, Michael A. and Garavelli, Marco and Olivucci,
  Massimo and Bernardi, Fernando}]{Robb_2007}
Robb MA, Garavelli M, Olivucci M, Bernardi F.
\newblock In: Lipkowitz KB, Boyd DB, editors. A {{Computational Strategy}} for
  {{Organic Photochemistry}} Hoboken, NJ, USA: {John Wiley \& Sons, Inc.};
  2007. p. 87--146.

\bibitem[{Navizet et~al.(2011)Isabelle Navizet and Ya-Jun Liu and Nicolas Ferre
  and Daniel {Roca-Sanjun} and Roland Lindh}]{Navizet_2011}
Navizet I, Liu YJ, Ferre N, {Roca-Sanjun} D, Lindh R.
\newblock The Chemistry of Bioluminescence: An Analysis of Chemical
  Functionalities.
\newblock ChemPhysChem 2011;12:3064--3076.

\bibitem[{Crespo-Otero and Barbatti(2018)Crespo-Otero, Rachel and Barbatti,
  Mario}]{Crespo_2018}
Crespo-Otero R, Barbatti M.
\newblock Recent Advances and Perspectives on Nonadiabatic Mixed
  Quantum--Classical Dynamics.
\newblock Chem Rev 2018;118:7026--7068.

\bibitem[{Robb(2018)Robb, Michael A}]{Robb_2018}
Robb MA.
\newblock Theoretical Chemistry for Electronic Excited States.
\newblock Theoretical and Computational Chemistry Series, The Royal Society of
  Chemistry; 2018.

\bibitem[{Mai and Gonz{\'a}lez(2020)Mai, Sebastian and Gonz{\'a}lez,
  Leticia}]{Mai_2020}
Mai S, Gonz{\'a}lez L.
\newblock Molecular Photochemistry: Recent Developments in Theory.
\newblock Angew Chem Int Ed 2020;59:16832--16846.

\bibitem[{Monkhorst(1977)Monkhorst, Hendrik J.}]{Monkhorst_1977}
Monkhorst HJ.
\newblock Calculation of properties with the coupled-cluster method.
\newblock International Journal of Quantum Chemistry 1977;12:421--432.

\bibitem[{Helgaker et~al.(1989)Helgaker, T. and J{\o}rgensen, P. and Handy,
  N.C.}]{Helgaker_1989}
Helgaker T, J{\o}rgensen P, Handy NC.
\newblock A Numerically Stable Procedure for Calculating M{\o}ller-Plesset
  Energy Derivatives, Derived Using the Theory of Lagrangians.
\newblock Theoret Chim Acta 1989;76:227--245.

\bibitem[{Koch et~al.(1990{\natexlab{a}})Koch, Henrik and Jensen, Hans Jo/rgen
  Aa. and Jo/rgensen, Poul and Helgaker, Trygve}]{Koch_1990}
Koch H, Jensen HJA, Jo/rgensen P, Helgaker T.
\newblock Excitation Energies from the Coupled Cluster Singles and Doubles
  Linear Response Function ({{CCSDLR}}). {{Applications}} to {{Be}}, {{CH}}
  {\textsuperscript{+}} , {{CO}}, and {{H}} {\textsubscript{2}} {{O}}.
\newblock J Chem Phys 1990;93:3345--3350.

\bibitem[{Koch et~al.(1990{\natexlab{b}})Koch, H. and Jensen, H. J. Aa. and
  Jorgensen, P. and Helgaker, T.}]{Koch_1990b}
Koch H, Jensen HJA, Jorgensen P, Helgaker T.
\newblock Excitation Energies from the Coupled Cluster Singles and Doubles
  Linear Response Function (CCSDLR). Applications to Be, CH$^+$, CO, and
  H$_2$O.
\newblock J Chem Phys 1990;93:3345--3350.

\bibitem[{Christiansen et~al.(1995)Christiansen, Ove and Koch, Henrik and
  J{\o}rgensen, Poul}]{Christiansen_1995b}
Christiansen O, Koch H, J{\o}rgensen P.
\newblock Response Functions in the CC3 Iterative Triple Excitation Model.
\newblock J Chem Phys 1995;103:7429--7441.

\bibitem[{Christiansen et~al.(1998)Christiansen, Ove and J{\o}rgensen, Poul and
  H\"attig, Christof}]{Christiansen_1998b}
Christiansen O, J{\o}rgensen P, H\"attig C.
\newblock Response Functions from Fourier Component Variational Perturbation
  Theory Applied to a Time-Averaged Quasienergy.
\newblock Int J Quantum Chem 1998;68:1--52.

\bibitem[{H{\"a}ttig(2003)H{\"a}ttig, Christof}]{Hattig_2003}
H{\"a}ttig C.
\newblock Geometry Optimizations with the Coupled-Cluster Model CC2 using the
  Resolution-of-the-Identity Approximation.
\newblock J Chem Phys 2003;118:7751--7761.

\bibitem[{K{\'a}llay and Gauss(2004)K{\'a}llay, Mih{\'a}ly and Gauss,
  J{\"u}rgen}]{Kallay_2004}
K{\'a}llay M, Gauss J.
\newblock Calculation of Excited-State Properties Using General Coupled-Cluster
  and Configuration-Interaction Models.
\newblock J Chem Phys 2004;121:9257--9269.

\bibitem[{H{\"a}ttig(2005)Christof H{\"a}ttig}]{Hattig_2005c}
H{\"a}ttig C.
\newblock Structure Optimizations for Excited States with Correlated
  Second-Order Methods: CC2 and ADC(2).
\newblock In: Jensen HJA, editor. Response Theory and Molecular Properties (A
  Tribute to Jan Linderberg and Poul J{\o}rgensen), vol.~50 of Advances in
  Quantum Chemistry Academic Press; 2005.p. 37--60.

\bibitem[{Dreuw and Wormit(2015)Dreuw, Andreas and Wormit,
  Michael}]{Dreuw_2015}
Dreuw A, Wormit M.
\newblock The Algebraic Diagrammatic Construction Scheme for the Polarization
  Propagator for the Calculation of Excited States.
\newblock WIREs Comput Mol Sci 2015;5:82--95.

\bibitem[{Pople et~al.(1989)Pople,John A. and Head?Gordon,Martin and
  Fox,Douglas J. and Raghavachari,Krishnan and Curtiss,Larry A.}]{Pople_1989}
Pople JA, Head?Gordon M, Fox DJ, Raghavachari K, Curtiss LA.
\newblock Gaussian?1 theory: A general procedure for prediction of molecular
  energies.
\newblock J Chem Phys 1989;90:5622--5629.

\bibitem[{Curtiss et~al.(1991)Curtiss, Larry A. and Raghavachari, Krishnan and
  Trucks, Gary W. and Pople, John A.}]{Curtiss_1991}
Curtiss LA, Raghavachari K, Trucks GW, Pople JA.
\newblock Gaussian-2 Theory for Molecular Energies of First- and Second-row
  Compounds.
\newblock J Chem Phys 1991;94:7221--7230.

\bibitem[{Curtiss et~al.(1997)Curtiss,Larry A. and Raghavachari,Krishnan and
  Redfern,Paul C. and Pople,John A.}]{Curtiss_1997}
Curtiss LA, Raghavachari K, Redfern PC, Pople JA.
\newblock Assessment of Gaussian-2 and density functional theories for the
  computation of enthalpies of formation.
\newblock J Chem Phys 1997;106:1063--1079.

\bibitem[{Curtiss et~al.(1998)Curtiss,Larry A. and Raghavachari,Krishnan and
  Redfern,Paul C. and Rassolov,Vitaly and Pople,John A.}]{Curtiss_1998}
Curtiss LA, Raghavachari K, Redfern PC, Rassolov V, Pople JA.
\newblock Gaussian-3 (G3) theory for molecules containing first and second-row
  atoms.
\newblock J Chem Phys 1998;109:7764--7776.

\bibitem[{Curtiss et~al.(2007)Curtiss,Larry A. and Redfern,Paul C. and
  Raghavachari,Krishnan}]{Curtiss_2007}
Curtiss LA, Redfern PC, Raghavachari K.
\newblock Gaussian-4 theory.
\newblock J Chem Phys 2007;126:084108.

\bibitem[{{\'A}ngy{\'a}n et~al.(2020){\'A}ngy{\'a}n, J{\'a}nos and Dobson, John
  and Jansen, Georg and Gould, Tim}]{Angyan_2020}
{\'A}ngy{\'a}n J, Dobson J, Jansen G, Gould T.
\newblock London Dispersion Forces in Molecules{,} Solids and Nano-structures.
\newblock Theoretical and Computational Chemistry Series, The Royal Society of
  Chemistry; 2020.

\bibitem[{Jure{\v c}ka et~al.(2006)Jure{\v c}ka, Petr and {\v S}poner, Ji{\v
  r}{\'\i} and {\v C}ern{\'y}, Ji{\v r}{\'\i} and Hobza, Pavel}]{Jureka_2006}
Jure{\v c}ka P, {\v S}poner J, {\v C}ern{\'y} J, Hobza P.
\newblock Benchmark database of accurate (MP2 and CCSD(T) complete basis set
  limit) interaction energies of small model complexes{,} DNA base pairs{,} and
  amino acid pairs.
\newblock Phys Chem Chem Phys 2006;8:1985--1993.

\bibitem[{{\v R}ez{\'a}{\v c} et~al.(2011){\v R}ez{\'a}{\v c}, Jan and Riley,
  Kevin E. and Hobza, Pavel}]{Rezac_2011}
{\v R}ez{\'a}{\v c} J, Riley KE, Hobza P.
\newblock S66: A Well-balanced Database of Benchmark Interaction Energies
  Relevant to Biomolecular Structures.
\newblock Journal of Chemical Theory and Computation 2011;7:2427--2438.
\newblock PMID: 21836824.

\bibitem[{{van Setten} et~al.(2015){van Setten}, Michiel J. and Caruso, Fabio
  and Sharifzadeh, Sahar and Ren, Xinguo and Scheffler, Matthias and Liu, Fang
  and Lischner, Johannes and Lin, Lin and Deslippe, Jack R. and Louie, Steven
  G. and Yang, Chao and Weigend, Florian and Neaton, Jeffrey B. and Evers,
  Ferdinand and Rinke, Patrick}]{vanSetten_2015}
{van Setten} MJ, Caruso F, Sharifzadeh S, Ren X, Scheffler M, Liu F, et~al.
\newblock {{{\emph{GW}}}} 100: {{Benchmarking}}
  {{{\emph{G}}}}{\textsubscript{0}}{{{\emph{W}}}}{\textsubscript{0}} for
  {{Molecular Systems}}.
\newblock J Chem Theory Comput 2015;11:5665--5687.

\bibitem[{Krause et~al.(2015)K. Krause and M. E. Harding and W.
  Klopper}]{Krause_2015}
Krause K, Harding ME, Klopper W.
\newblock Coupled-Cluster Reference Values For The Gw27 And Gw100 Test Sets For
  The Assessment Of Gw Methods.
\newblock Mol Phys 2015;113:1952.

\bibitem[{Maggio and Kresse(2016)Maggio, Emanuele and Kresse,
  Georg}]{Maggio_2016}
Maggio E, Kresse G.
\newblock Correlation energy for the homogeneous electron gas: Exact
  Bethe-Salpeter solution and an approximate evaluation.
\newblock Phys Rev B 2016;93:235113.

\bibitem[{Stuke et~al.(2020)Annika Stuke and Christian Kunkel and Dorothea
  Golze and Milica Todorovi{\'c} and Johannes T. Margraf and Karsten Reuter and
  Patrick Rinke and Harald Oberhofer}]{Stuke_2020}
Stuke A, Kunkel C, Golze D, Todorovi{\'c} M, Margraf JT, Reuter K, et~al.
\newblock Atomic Structures and Orbital Energies of 61,489 Crystal-Forming
  Organic Molecules.
\newblock Sci Data 2020;7:58.

\bibitem[{{van Setten} et~al.(2013){van Setten}, M. J. and Weigend, F. and
  Evers, F.}]{vanSetten_2013}
{van Setten} MJ, Weigend F, Evers F.
\newblock The {{{\emph{GW}}}} -{{Method}} for {{Quantum Chemistry
  Applications}}: {{Theory}} and {{Implementation}}.
\newblock J Chem Theory Comput 2013;9:232--246.

\bibitem[{Bruneval et~al.(2016)Bruneval, Fabien and Rangel, Tonatiuh and Hamed,
  Samia M. and Shao, Meiyue and Yang, Chao and Neaton, Jeffrey
  B.}]{Bruneval_2016}
Bruneval F, Rangel T, Hamed SM, Shao M, Yang C, Neaton JB.
\newblock Molgw 1: {{Many}}-Body Perturbation Theory Software for Atoms,
  Molecules, and Clusters.
\newblock Comput Phys Commun 2016;208:149--161.

\bibitem[{Caruso et~al.(2016)F. Caruso and M. Dauth and M. J. {van Setten} and
  P. Rinke}]{Caruso_2016}
Caruso F, Dauth M, {van Setten} MJ, Rinke P.
\newblock Benchmark of GW Approaches for the GW100 Test Set.
\newblock J Chem Theory Comput 2016;12:5076.

\bibitem[{Govoni and Galli(2018)Marco Govoni and Giulia Galli}]{Govoni_2018}
Govoni M, Galli G.
\newblock GW100: Comparison of Methods and Accuracy of Results Obtained with
  the WEST Code.
\newblock J Chem Theory Comput 2018;14:1895--1909.

\bibitem[{Tajti et~al.(2004)Tajti,Attila and Szalay,P{\'e}ter G. and
  Cs{\'a}sz{\'a}r,Attila G. and K{\'a}llay,Mih{\'a}ly and Gauss,J{\"u}rgen and
  Valeev,Edward F. and Flowers,Bradley A. and V{\'a}zquez,Juana and
  Stanton,John F.}]{Tajti_2004}
Tajti A, Szalay PG, Cs{\'a}sz{\'a}r AG, K{\'a}llay M, Gauss J, Valeev EF,
  et~al.
\newblock HEAT: High accuracy extrapolated ab initio thermochemistry.
\newblock J Chem Phys 2004;121:11599--11613.

\bibitem[{Bomble et~al.(2006)Bomble,Yannick J. and V{\'a}zquez,Juana and
  K{\'a}llay,Mih{\'a}ly and Michauk,Christine and Szalay,P{\'e}ter G. and
  Cs{\'a}sz{\'a}r,Attila G. and Gauss,J{\"u}rgen and Stanton,John
  F.}]{Bomble_2006}
Bomble YJ, V{\'a}zquez J, K{\'a}llay M, Michauk C, Szalay PG, Cs{\'a}sz{\'a}r
  AG, et~al.
\newblock High-accuracy extrapolated ab initio thermochemistry. II. Minor
  improvements to the protocol and a vital simplification.
\newblock J Chem Phys 2006;125:064108.

\bibitem[{Harding et~al.(2008)M. E. Harding and J. Vazquez and B. Ruscic and A.
  K. Wilson and J. Gauss and J. F. Stanton}]{Harding_2008}
Harding ME, Vazquez J, Ruscic B, Wilson AK, Gauss J, Stanton JF.
\newblock High-Accuracy Extrapolated ab Initio Thermochemistry. III. Additional
  Improvements and Overview.
\newblock J Chem Phys 2008;128:114111.

\bibitem[{Motta et~al.(2017)Motta, Mario and Ceperley, David M and Chan, Garnet
  Kin-Lic and Gomez, John A and Gull, Emanuel and Guo, Sheng and
  Jim{\'e}nez-Hoyos, Carlos A and Lan, Tran Nguyen and Li, Jia and Ma, Fengjie
  and others}]{Motta_2017}
Motta M, Ceperley DM, Chan GKL, Gomez JA, Gull E, Guo S, et~al.
\newblock Towards the solution of the many-electron problem in real materials:
  Equation of state of the hydrogen chain with state-of-the-art many-body
  methods.
\newblock Phys Rev X 2017;7:031059.

\bibitem[{Williams et~al.(2020)Williams, Kiel T and Yao, Yuan and Li, Jia and
  Chen, Li and Shi, Hao and Motta, Mario and Niu, Chunyao and Ray, Ushnish and
  Guo, Sheng and Anderson, Robert J and others}]{Williams_2020}
Williams KT, Yao Y, Li J, Chen L, Shi H, Motta M, et~al.
\newblock Direct comparison of many-body methods for realistic electronic
  Hamiltonians.
\newblock Phys Rev X 2020;10:011041.

\bibitem[{Parr and Yang(1989)R. G. Parr and W. Yang}]{ParrBook}
Parr RG, Yang W.
\newblock Density-Functional Theory of Atoms and Molecules.
\newblock Clarendon Press: Oxford; 1989.

\bibitem[{Zhao and Truhlar(2006)Zhao,Yan and Truhlar,Donald G.}]{Zhao_2006}
Zhao Y, Truhlar DG.
\newblock Comparative assessment of density functional methods for 3d
  transition-metal chemistry.
\newblock J Chem Phys 2006;124:224105.

\bibitem[{Mardirossian and Head-Gordon(2017)Narbe Mardirossian and Martin
  Head-Gordon}]{Mardirossian_2017}
Mardirossian N, Head-Gordon M.
\newblock Thirty years of density functional theory in computational chemistry:
  an overview and extensive assessment of 200 density functionals.
\newblock Mol Phys 2017;115(19):2315--2372.

\bibitem[{Goerigk and Grimme(2010)Goerigk, Lars and Grimme,
  Stefan}]{Goerigk_2010}
Goerigk L, Grimme S.
\newblock A General Database for Main Group Thermochemistry, Kinetics, and
  Noncovalent Interactions -- Assessment of Common and Reparameterized
  (meta-)GGA Density Functionals.
\newblock J Chem Theory Comput 2010;6:107--126.
\newblock PMID: 26614324.

\bibitem[{Goerigk and Grimme(2011{\natexlab{a}})Goerigk, Lars and Grimme,
  Stefan}]{Goerigk_2011a}
Goerigk L, Grimme S.
\newblock A thorough benchmark of density functional methods for general main
  group thermochemistry{,} kinetics{,} and noncovalent interactions.
\newblock Phys Chem Chem Phys 2011;13:6670--6688.

\bibitem[{Goerigk and Grimme(2011{\natexlab{b}})Goerigk, Lars and Grimme,
  Stefan}]{Goerigk_2011b}
Goerigk L, Grimme S.
\newblock Efficient and Accurate Double-Hybrid-Meta-GGA Density
  Functionals---Evaluation with the Extended GMTKN30 Database for General Main
  Group Thermochemistry, Kinetics, and Noncovalent Interactions.
\newblock J Chem Theory Comput 2011;7:291--309.
\newblock PMID: 26596152.

\bibitem[{Goerigk et~al.(2017)Goerigk, Lars and Hansen, Andreas and Bauer,
  Christoph and Ehrlich, Stephan and Najibi, Asim and Grimme,
  Stefan}]{Goerigk_2017}
Goerigk L, Hansen A, Bauer C, Ehrlich S, Najibi A, Grimme S.
\newblock A look at the density functional theory zoo with the advanced GMTKN55
  database for general main group thermochemistry{,} kinetics and noncovalent
  interactions.
\newblock Phys Chem Chem Phys 2017;19:32184--32215.

\bibitem[{Schreiber et~al.(2008)Schreiber, M. and Silva-Junior, M. R. and
  Sauer, S. P. A. and Thiel, W.}]{Schreiber_2008}
Schreiber M, Silva-Junior MR, Sauer SPA, Thiel W.
\newblock Benchmarks for Electronically Excited States: CASPT2, CC2, CCSD and
  CC3.
\newblock J Chem Phys 2008;128:134110.

\bibitem[{Silva-Junior et~al.(2008)Silva-Junior, M. R. and Schreiber, M. and
  Sauer, S. P. A. and Thiel, W.}]{Silva-Junior_2008}
Silva-Junior MR, Schreiber M, Sauer SPA, Thiel W.
\newblock Benchmarks for Electronically Excited States: Time-Dependent Density
  Functional Theory and Density Functional Theory Based Multireference
  Configuration Interaction.
\newblock J Chem Phys 2008;129:104103.

\bibitem[{Silva-Junior et~al.(2010{\natexlab{a}})Silva-Junior, Mario R. and
  Schreiber, Marko and Sauer, Stephan P. A. and Thiel,
  Walter}]{Silva-Junior_2010}
Silva-Junior MR, Schreiber M, Sauer SPA, Thiel W.
\newblock Benchmarks of Electronically Excited States: {{Basis}} Set Effects on
  {{CASPT2}} Results.
\newblock J Chem Phys 2010;133:174318.

\bibitem[{Silva-Junior et~al.(2010{\natexlab{b}})Silva-Junior, Mario R. and
  Sauer, Stephan P.A. and Schreiber, Marko and Thiel,
  Walter}]{Silva-Junior_2010b}
Silva-Junior MR, Sauer SPA, Schreiber M, Thiel W.
\newblock Basis Set Effects on Coupled Cluster Benchmarks of Electronically
  Excited States: {{CC3}}, {{CCSDR}}(3) and {{CC2}}.
\newblock Mol Phys 2010;108:453--465.

\bibitem[{Silva-Junior et~al.(2010{\natexlab{c}})Silva-Junior, M. R. and
  Schreiber, M. and Sauer, S. P. A. and Thiel, W.}]{Silva-Junior_2010c}
Silva-Junior MR, Schreiber M, Sauer SPA, Thiel W.
\newblock Benchmarks of Electronically Excited States: Basis Set Effecs
  Benchmarks of Electronically Excited States: Basis Set Effects on CASPT2
  Results.
\newblock J Chem Phys 2010;133:174318.

\bibitem[{Christiansen et~al.(1995)Ove Christiansen and Henrik Koch and Poul
  J{\o}rgensen}]{Christiansen_1995a}
Christiansen O, Koch H, J{\o}rgensen P.
\newblock The Second-Order Approximate Coupled Cluster Singles and Doubles
  Model CC2.
\newblock Chem Phys Lett 1995;243:409--418.

\bibitem[{H\"attig and Weigend(2000)H\"attig, C. and Weigend, F.}]{Hattig_2000}
H\"attig C, Weigend F.
\newblock CC2 Excitation Energy Calculations on Large Molecules Using the
  Resolution of the Identity Approximation.
\newblock J Chem Phys 2000;113:5154--5161.

\bibitem[{ROWE(1968)ROWE, D. J.}]{Rowe_1968}
ROWE DJ.
\newblock Equations-of-Motion Method and the Extended Shell Model.
\newblock Rev Mod Phys 1968;40:153--166.

\bibitem[{Stanton and Bartlett(1993)Stanton,John F. and Bartlett,Rodney
  J.}]{Stanton_1993}
Stanton JF, Bartlett RJ.
\newblock The equation of motion coupled?cluster method. A systematic
  biorthogonal approach to molecular excitation energies, transition
  probabilities, and excited state properties.
\newblock J Chem Phys 1993;98:7029--7039.

\bibitem[{Koch et~al.(1994)Koch,Henrik and Kobayashi,Rika and Sanchez de
  Mer{\'a}s,Alfredo and Jo/rgensen,Poul}]{Koch_1994}
Koch H, Kobayashi R, Sanchez~de Mer{\'a}s A, Jo/rgensen P.
\newblock Calculation of size?intensive transition moments from the coupled
  cluster singles and doubles linear response function.
\newblock J Chem Phys 1994;100:4393--4400.

\bibitem[{Koch et~al.(1997)Koch, Henrik and Christiansen, Ove and Jorgensen,
  Poul and Sanchez de Mer{\'a}s, Alfredo M. and Helgaker, Trygve}]{Koch_1997}
Koch H, Christiansen O, Jorgensen P, Sanchez~de Mer{\'a}s AM, Helgaker T.
\newblock The CC3 Model: An Iterative Coupled Cluster Approach Including
  Connected Triples.
\newblock J Chem Phys 1997;106:1808--1818.

\bibitem[{Andersson et~al.(1990)Andersson, Kerstin. and Malmqvist, Per Aake.
  and Roos, Bjoern O. and Sadlej, Andrzej J. and Wolinski,
  Krzysztof.}]{Andersson_1990}
Andersson K, Malmqvist PA, Roos BO, Sadlej AJ, Wolinski K.
\newblock Second-Order Perturbation Theory With a CASSCF Reference Function.
\newblock J Phys Chem 1990;94:5483--5488.

\bibitem[{Andersson et~al.(1992)Andersson, Kerstin and Malmqvist, Per-Ake and
  Roos, Bj{\"o}rn O.}]{Andersson_1992}
Andersson K, Malmqvist PA, Roos BO.
\newblock Second-Order Perturbation Theory With a Complete Active Space
  Self-Consistent Field Reference Function.
\newblock J Chem Phys 1992;96:1218--1226.

\bibitem[{B.~O.~Roos et~al.(1996)B. O. Roos, K. Andersson and M. P. Fulscher
  and P.-A. Malmqvist and L. {Serrano-Andres}}]{Roos}
B~O~Roos KA, Fulscher MP, Malmqvist PA, {Serrano-Andres} L.
\newblock In: Prigogine I, Rice SA, editors. Adv. Chem. Phys., vol. XCIII
  Wiley, New York; 1996. p. 219--331.

\bibitem[{Leang et~al.(2012)Leang,Sarom S. and Zahariev,Federico and
  Gordon,Mark S.}]{Leang_2012}
Leang SS, Zahariev F, Gordon MS.
\newblock Benchmarking the performance of time-dependent density functional
  methods.
\newblock J Chem Phys 2012;136:104101.

\bibitem[{Runge and Gross(1984)Runge, E. and Gross, E. K. U.}]{Runge_1984}
Runge E, Gross EKU.
\newblock Density-Functional Theory for Time-Dependent Systems.
\newblock Phys Rev Lett 1984;52:997--1000.

\bibitem[{Casida(1995)M. E. Casida}]{Casida_1995}
Casida ME.
\newblock In: Chong DP, editor. Time-Dependent Density Functional Response
  Theory for Molecules Recent Advances in Density Functional Methods, World
  Scientific, Singapore; 1995. p. 155--192.

\bibitem[{Casida and Huix-Rotllant(2012)Casida, M.E. and Huix-Rotllant,
  M.}]{Casida_2012}
Casida ME, Huix-Rotllant M.
\newblock Progress in Time-Dependent Density-Functional Theory.
\newblock Annu Rev Phys Chem 2012;63:287.

\bibitem[{Ullrich(2012)Ullrich, C.}]{Ulrich_2012}
Ullrich C.
\newblock Time-Dependent Density-Functional Theory: Concepts and Applications.
\newblock Oxford Graduate Texts, New York: Oxford University Press; 2012.

\bibitem[{Schwabe and Goerigk(????)T. Schwabe and L. Goerigk}]{Schwabe_2017}
Schwabe T, Goerigk L.
\newblock Time-Dependent Double-Hybrid Density Functionals with Spin-Component
  and Spin-Opposite Scaling.
\newblock J Chem Theory Comput;13:4307.

\bibitem[{Casanova-Paez et~al.(2019)M. Casanova-Paez and M. B. Dardis and L.
  Goerigk}]{Casanova-Paez_2019}
Casanova-Paez M, Dardis MB, Goerigk L.
\newblock $\omega$B2PLYP and $\omega$B2GPPLYP: The First Two Double-Hybrid
  Density Functionals with Long-Range Correction Optimized for Excitation
  Energies.
\newblock J Chem Theory Comput 2019;15:4735.

\bibitem[{Casanova-Paez and Goerigk(2020)M. Casanova-Paez and L.
  Goerigk}]{Casanova_Paes_2020}
Casanova-Paez M, Goerigk L.
\newblock Assessing the Tamm--Dancoff approximation, singlet--singlet, and
  singlet--triplet excitations with the latest long-range corrected
  double-hybrid density functionals.
\newblock J Chem Phys 2020;153:064106.

\bibitem[{Furche and Ahlrichs(2002)F. Furche and R. Ahlrichs}]{Furche_2002}
Furche F, Ahlrichs R.
\newblock Adiabatic Time-Dependent Density Functional Methods for Excited State
  Properties.
\newblock J Chem Phys 2002;117:7433.

\bibitem[{Send et~al.(2011)Send, Robert and K{\"u}hn, Michael and Furche,
  Filipp}]{Send_2011a}
Send R, K{\"u}hn M, Furche F.
\newblock Assessing Excited State Methods by Adiabatic Excitation Energies.
\newblock J Chem Theory Comput 2011;7:2376--2386.

\bibitem[{Winter et~al.(2013)Winter, Nina O. C. and Graf, Nora K. and
  Leutwyler, Samuel and H{\"a}ttig, Christof}]{Winter_2013}
Winter NOC, Graf NK, Leutwyler S, H{\"a}ttig C.
\newblock Benchmarks for 0--0 Transitions of Aromatic Organic Molecules:
  DFT/B3LYP{,} ADC(2){,} CC2{,} SOS-CC2 and SCS-CC2 Compared to High-resolution
  Gas-Phase Data.
\newblock Phys Chem Chem Phys 2013;15:6623--6630.

\bibitem[{Loos et~al.(2018)Loos, Pierre-Fran{\c c}ois and Galland, Nicolas and
  Jacquemin, Denis}]{Loos_2018}
Loos PF, Galland N, Jacquemin D.
\newblock Theoretical 0--0 Energies with Chemical Accuracy.
\newblock J Phys Chem Lett 2018;9:4646--4651.

\bibitem[{Loos and Jacquemin(2019{\natexlab{a}})P. F. Loos and D.
  Jacquemin}]{Loos_2019a}
Loos PF, Jacquemin D.
\newblock Chemically Accurate 0-0 Energies With not-so-Accurate Excited State
  Geometries.
\newblock J Chem Theory Comput 2019;15:2481.

\bibitem[{Loos and Jacquemin(2019{\natexlab{b}})Loos, Pierre-Francois and
  Jacquemin, Denis}]{Loos_2019b}
Loos PF, Jacquemin D.
\newblock Evaluating 0-0 Energies with Theoretical Tools: a Short Review.
\newblock ChemPhotoChem 2019;3:684--696.

\bibitem[{Dierksen and Grimme(2004)Dierksen, M. and Grimme, S.}]{Dierksen_2004}
Dierksen M, Grimme S.
\newblock A density functional calculation of the vibronic structure of
  electronic absorption spectra.
\newblock J Chem Phys 2004;120:3544--3554.

\bibitem[{Goerigk and Grimme(2010)Goerigk, L. and Grimme, S.}]{Goerigk_2010a}
Goerigk L, Grimme S.
\newblock Assessment of TD-DFT Methods and of Various Spin Scaled CIS$_n$D and
  CC2 Versions for the Treatment of Low-Lying Valence Excitations of Large
  Organic Dyes.
\newblock J Chem Phys 2010;132:184103.

\bibitem[{Jacquemin et~al.(2012)Jacquemin , Denis and Planchat, Aur{\'e}lien
  and Adamo, Carlo and Mennucci, Benedetta}]{Jacquemin_2012}
Jacquemin D, Planchat A, Adamo C, Mennucci B.
\newblock A TD-DFT Assessment of Functionals for Optical 0-0 Transitions in
  Solvated Dyes.
\newblock J Chem Theory Comput 2012;8:2359--2372.

\bibitem[{Jacquemin et~al.(2015)Jacquemin, Denis and Duchemin, Ivan and Blase,
  Xavier}]{Jacquemin_2015b}
Jacquemin D, Duchemin I, Blase X.
\newblock 0--0 Energies Using Hybrid Schemes: Benchmarks of TD-DFT, CIS(D),
  ADC(2), CC2, and BSE/GW formalisms for 80 Real-Life Compounds.
\newblock J Chem Theory Comput 2015;11:5340--5359.

\bibitem[{Kozma et~al.(2020)B. Kozma and A. Tajti and B. Demoulin and R. Izsak
  and M. Nooijen and P. G. Szalay}]{Kozma_2020}
Kozma B, Tajti A, Demoulin B, Izsak R, Nooijen M, Szalay PG.
\newblock A New Benchmark Set for Excitation Energy of Charge Transfer States:
  Systematic Investigation of Coupled Cluster Type Methods.
\newblock J Chem Theory Comput 2020;16:4213--4225.

\bibitem[{Hoyer et~al.(2016)Hoyer, Chad E. and Ghosh, Soumen and Truhlar,
  Donald G. and Gagliardi, Laura}]{Hoyer_2016}
Hoyer CE, Ghosh S, Truhlar DG, Gagliardi L.
\newblock Multiconfiguration Pair-Density Functional Theory Is as Accurate as
  CASPT2 for Electronic Excitation.
\newblock J Phys Chem Lett 2016;7:586--591.

\bibitem[{Loos et~al.(2018)P. F. Loos and A. Scemama and A. Blondel and Y.
  Garniron and M. Caffarel and D. Jacquemin}]{Loos_2018a}
Loos PF, Scemama A, Blondel A, Garniron Y, Caffarel M, Jacquemin D.
\newblock A Mountaineering Strategy to Excited States: Highly-Accurate
  Reference Energies and Benchmarks.
\newblock J Chem Theory Comput 2018;14:4360.

\bibitem[{Loos et~al.(2019)Loos, Pierre-Fran{\c c}ois and Boggio-Pasqua,
  Martial and Scemama, Anthony and Caffarel, Michel and Jacquemin,
  Denis}]{Loos_2019}
Loos PF, Boggio-Pasqua M, Scemama A, Caffarel M, Jacquemin D.
\newblock Reference Energies for Double Excitations.
\newblock J Chem Theory Comput 2019;15:1939--1956.

\bibitem[{Loos et~al.(2020{\natexlab{a}})P. F. Loos and F. Lipparini and M.
  Boggio-Pasqua and A. Scemama and D. Jacquemin}]{Loos_2020b}
Loos PF, Lipparini F, Boggio-Pasqua M, Scemama A, Jacquemin D.
\newblock A Mountaineering Strategy to Excited States: Highly-Accurate Energies
  and Benchmarks for Medium Size Molecules,.
\newblock J Chem Theory Comput 2020;16:1711.

\bibitem[{Loos et~al.(2020{\natexlab{b}})P. F. Loos and A. Scemama and M.
  Boggio-Pasqua and D. Jacquemin}]{Loos_2020c}
Loos PF, Scemama A, Boggio-Pasqua M, Jacquemin D.
\newblock A Mountaineering Strategy to Excited States: Highly-Accurate Energies
  and Benchmarks for Exotic Molecules and Radicals.
\newblock J Chem Theory Comput 2020;16:3720--3736.

\bibitem[{K{\"o}hn and H{\"a}ttig(2003)K{\"o}hn, Andreas and H{\"a}ttig,
  Christof}]{Kohn_2003}
K{\"o}hn A, H{\"a}ttig C.
\newblock Analytic Gradients for Excited States in the Coupled-Cluster Model
  CC2 Employing the Resolution-Of-The-Identity Approximation.
\newblock J Chem Phys 2003;119:5021--5036.

\bibitem[{Fang et~al.(2014)Fang, Changfeng and Oruganti, Baswanth and Durbeej,
  Bo}]{Fang_2014}
Fang C, Oruganti B, Durbeej B.
\newblock How Method-Dependent Are Calculated Differences Between Vertical,
  Adiabatic and 0-0 Excitation Energies?
\newblock J Phys Chem A 2014;118:4157--4171.

\bibitem[{Oruganti et~al.(2016)Baswanth Oruganti and Changfeng Fang and Bo
  Durbeej}]{Oruganti_2016}
Oruganti B, Fang C, Durbeej B.
\newblock Assessment of a Composite CC2/DFT Procedure for Calculating 0--0
  Excitation Energies of Organic Molecules.
\newblock Mol Phys 2016;114:3448--3463.

\bibitem[{Noga and Bartlett(1987)Jozef Noga and Rodney J. Bartlett}]{Noga_1987}
Noga J, Bartlett RJ.
\newblock The Full CCSDT Model for Molecular Electronic Structure.
\newblock J Chem Phys 1987;86:7041--7050.

\bibitem[{Kucharski and Bartlett(1991)Kucharski, Stanislaw A. and Bartlett,
  Rodney J.}]{Kucharski_1991}
Kucharski SA, Bartlett RJ.
\newblock Recursive Intermediate Factorization and Complete Computational
  Linearization of the Coupled-Cluster Single, Double, Triple, and Quadruple
  Excitation Equations.
\newblock Theor Chim Acta 1991;80:387--405.

\bibitem[{Kucharski et~al.(2001)Stanis{\l}aw A. Kucharski and Marta W{\l}och
  and Monika Musia{\l} and Rodney J. Bartlett}]{Kucharski_2001}
Kucharski SA, W{\l}och M, Musia{\l} M, Bartlett RJ.
\newblock Coupled-Cluster Theory for Excited Electronic States: The Full
  Equation-Of-Motion Coupled-Cluster Single, Double, and Triple Excitation
  Method.
\newblock J Chem Phys 2001;115:8263--8266.

\bibitem[{Kowalski and Piecuch(2001)Kowalski, K. and Piecuch,
  P.}]{Kowalski_2001}
Kowalski K, Piecuch P.
\newblock The Active-Space Equation-of-Motion Coupled-Cluster Methods for
  Excited Electronic States: Full EOMCCSDt.
\newblock J Chem Phys 2001;115:643--651.

\bibitem[{K{\'a}llay et~al.(2003)K{\'a}llay,Mih{\'a}ly and Gauss,J{\"u}rgen and
  Szalay,P{\'e}ter G.}]{Kallay_2003}
K{\'a}llay M, Gauss J, Szalay PG.
\newblock Analytic First Derivatives for General Coupled-Cluster and
  Configuration Interaction Models.
\newblock J Chem Phys 2003;119:2991--3004.

\bibitem[{Hirata et~al.(2000)Hirata, So and Nooijen, Marcel and Bartlett,
  Rodney J.}]{Hirata_2000}
Hirata S, Nooijen M, Bartlett RJ.
\newblock High-Order Determinantal Equation-of-Motion Coupled-Cluster
  Calculations for Electronic Excited States.
\newblock Chem Phys Lett 2000;326:255--262.

\bibitem[{Hirata(2004)Hirata, S.}]{Hirata_2004}
Hirata S.
\newblock Higher-Order Equation-of-Motion Coupled-Cluster Methods.
\newblock J Chem Phys 2004;121:51--59.

\bibitem[{Booth et~al.(2009)Booth, George H. and Thom, Alex J. W. and Alavi,
  Ali}]{Booth_2009}
Booth GH, Thom AJW, Alavi A.
\newblock Fermion {Monte} {Carlo} without fixed nodes: {A} game of life, death,
  and annihilation in {Slater} determinant space.
\newblock J Chem Phys 2009;131:054106.

\bibitem[{Booth and Alavi(2010)George H. Booth and Ali Alavi}]{Booth_2010}
Booth GH, Alavi A.
\newblock Approaching chemical accuracy using full configuration-interaction
  quantum Monte Carlo: A study of ionization potentials.
\newblock J Chem Phys 2010;132:174104.

\bibitem[{Cleland et~al.(2010)Deidre Cleland and George H. Booth and Ali
  Alavi}]{Cleland_2010}
Cleland D, Booth GH, Alavi A.
\newblock Communications: Survival of the fittest: Accelerating convergence in
  full configuration-interaction quantum Monte Carlo.
\newblock J Chem Phys 2010;132:041103.

\bibitem[{Booth et~al.(2011)Booth, George H. and Cleland, Deidre and Thom, Alex
  J. W. and Alavi, Ali}]{Booth_2011}
Booth GH, Cleland D, Thom AJW, Alavi A.
\newblock Breaking the Carbon Dimer: {{The}} Challenges of Multiple Bond
  Dissociation with Full Configuration Interaction Quantum {{Monte Carlo}}
  Methods.
\newblock J Chem Phys 2011;135:084104.

\bibitem[{Daday et~al.(2012)Daday, Csaba and Smart, Simon and Booth, George H.
  and Alavi, Ali and Filippi, Claudia}]{Daday_2012}
Daday C, Smart S, Booth GH, Alavi A, Filippi C.
\newblock Full {{Configuration Interaction Excitations}} of {{Ethene}} and
  {{Butadiene}}: {{Resolution}} of an {{Ancient Question}}.
\newblock J Chem Theory Comput 2012;8:4441--4451.

\bibitem[{Blunt et~al.(2015)Blunt, N. S. and Smart, Simon D. and Booth, George
  H. and Alavi, Ali}]{Blunt_2015}
Blunt NS, Smart SD, Booth GH, Alavi A.
\newblock An Excited-State Approach within Full Configuration Interaction
  Quantum {{Monte Carlo}}.
\newblock J Chem Phys 2015;143:134117.

\bibitem[{Ghanem et~al.(2019)Ghanem, K. and Lozovoi, A. Y. and Alavi,
  A.}]{Ghanem_2019}
Ghanem K, Lozovoi AY, Alavi A.
\newblock Unbiasing the Initiator Approximation in Full Configuration
  Interaction Quantum Monte Carlo.
\newblock J Chem Phys 2019;151:224108.

\bibitem[{Deustua et~al.(2017)Deustua, J. Emiliano and Shen, Jun and Piecuch,
  Piotr}]{Deustua_2017}
Deustua JE, Shen J, Piecuch P.
\newblock Converging High-Level Coupled-Cluster Energetics by Monte Carlo
  Sampling and Moment Expansions.
\newblock Phys Rev Lett 2017;119:223003.

\bibitem[{Deustua et~al.(2018)Deustua, J. E. and Magoulas, I. and Shen, J. and
  Piecuch, P.}]{Deustua_2018}
Deustua JE, Magoulas I, Shen J, Piecuch P.
\newblock Communication: Approaching Exact Quantum Chemistry by Cluster
  Analysis of Full Configuration Interaction Quan- tum Monte Carlo Wave
  Functions.
\newblock J Chem Phys 2018;149:151101.

\bibitem[{Holmes et~al.(2017)Holmes, Adam A. and Umrigar, C. J. and Sharma,
  Sandeep}]{Holmes_2017}
Holmes AA, Umrigar CJ, Sharma S.
\newblock Excited states using semistochastic heat-bath configuration
  interaction.
\newblock J Chem Phys 2017;147:164111.

\bibitem[{Chien et~al.(2018)Chien, Alan D. and Holmes, Adam A. and Otten,
  Matthew and Umrigar, C. J. and Sharma, Sandeep and Zimmerman, Paul
  M.}]{Chien_2018}
Chien AD, Holmes AA, Otten M, Umrigar CJ, Sharma S, Zimmerman PM.
\newblock Excited {{States}} of {{Methylene}}, {{Polyenes}}, and {{Ozone}} from
  {{Heat}}-{{Bath Configuration Interaction}}.
\newblock J Phys Chem A 2018;122:2714--2722.

\bibitem[{Li et~al.(2018)J. Li and M. Otten and A. A. Holmes and S. Sharma and
  C. J. Umrigar}]{Li_2018}
Li J, Otten M, Holmes AA, Sharma S, Umrigar CJ.
\newblock Fast semistochastic heat-bath configuration interaction.
\newblock J Chem Phys 2018;149:214110.

\bibitem[{Yao et~al.(2020)Yuan Yao and Emmanuel Giner and Junhao Li and Julien
  Toulouse and C. J. Umrigar}]{Yao_2020}
Yao Y, Giner E, Li J, Toulouse J, Umrigar CJ, Almost exact energies for the
  Gaussian-2 set with the semistochastic heat-bath configuration interaction
  method; 2020.

\bibitem[{Li et~al.(2020)Li, Junhao and Yao, Yuan and Holmes, Adam A. and
  Otten, Matthew and Sun, Qiming and Sharma, Sandeep and Umrigar, C.
  J.}]{Li_2020}
Li J, Yao Y, Holmes AA, Otten M, Sun Q, Sharma S, et~al.
\newblock Accurate many-body electronic structure near the basis set limit:
  Application to the chromium dimer.
\newblock Phys Rev Research 2020;2:012015.

\bibitem[{Eriksen et~al.(2017)J. J. Eriksen and F. Lipparini and J.
  Gauss}]{Eriksen_2017}
Eriksen JJ, Lipparini F, Gauss J.
\newblock Virtual Orbital Many-Body Expansions: A Possible Route towards the
  Full Configuration Interaction Limit.
\newblock J Phys Chem Lett 2017;8:4633--4639.

\bibitem[{Eriksen and Gauss(2018)J. J. Eriksen and J. Gauss}]{Eriksen_2018}
Eriksen JJ, Gauss J.
\newblock Many-Body Expanded Full Configuration Interaction. I. Weakly
  Correlated Regime.
\newblock J Chem Theory Comput 2018;14:5180.

\bibitem[{Eriksen and Gauss(2019{\natexlab{a}})J. J. Eriksen and J.
  Gauss}]{Eriksen_2019a}
Eriksen JJ, Gauss J.
\newblock Many-Body Expanded Full Configuration Interaction. II. Strongly
  Correlated Regime.
\newblock J Chem Theory Comput 2019;15:4873.

\bibitem[{Eriksen and Gauss(2019{\natexlab{b}})J. J. Eriksen and J.
  Gauss}]{Eriksen_2019b}
Eriksen JJ, Gauss J.
\newblock Generalized Many-Body Expanded Full Configuration Interaction Theory.
\newblock J Phys Chem Lett 2019;27:7910--7915.

\bibitem[{Xu et~al.(2018)Xu, E. and Uejima, M. and Ten-no, S. L.}]{Xu_2018}
Xu E, Uejima M, Ten-no SL.
\newblock Full Coupled-Cluster Reduction for Accurate Description of Strong
  Electron Correlation.
\newblock Phys Rev Lett 2018;121:113001.

\bibitem[{Xu et~al.(2020)Enhua Xu and Motoyuki Uejima and Seiichiro L.
  Ten-no}]{Xu_2020}
Xu E, Uejima M, Ten-no SL, Towards near-exact solutions of molecular electronic
  structure: Full coupled-cluster reduction with a second-order perturbative
  correction; 2020.

\bibitem[{Loos et~al.(2020)Loos,Pierre-Fran{\c c}ois and Damour,Yann and
  Scemama,Anthony}]{Loos_2020e}
Loos PF, Damour Y, Scemama A.
\newblock The performance of CIPSI on the ground state electronic energy of
  benzene.
\newblock J Chem Phys 2020;153:176101.

\bibitem[{Eriksen(2020)Janus J. Eriksen}]{Eriksen_2021}
Eriksen JJ, The Shape of FCI to Come; 2020.

\bibitem[{Bender and Davidson(1969)Bender, Charles F. and Davidson, Ernest
  R.}]{Bender_1969}
Bender CF, Davidson ER.
\newblock Studies in Configuration Interaction: The First-Row Diatomic
  Hydrides.
\newblock Phys Rev 1969;183:23--30.

\bibitem[{Whitten and Hackmeyer(1969)Whitten, J. L. and Hackmeyer,
  Melvyn}]{Whitten_1969}
Whitten JL, Hackmeyer M.
\newblock Configuration Interaction Studies of Ground and Excited States of
  Polyatomic Molecules. I. The CI Formulation and Studies of Formaldehyde.
\newblock J Chem Phys 1969;51:5584--5596.

\bibitem[{Huron et~al.(1973)Huron, B. and Malrieu, J. P. and Rancurel,
  P.}]{Huron_1973}
Huron B, Malrieu JP, Rancurel P.
\newblock Iterative perturbation calculations of ground and excited state
  energies from multiconfigurational zeroth?order wavefunctions.
\newblock J Chem Phys 1973;58:5745--5759.

\bibitem[{Abrams and Sherrill(2005)Abrams, Micah L. and Sherrill, C.
  David}]{Abrams_2005}
Abrams ML, Sherrill CD.
\newblock Important configurations in configuration interaction and
  coupled-cluster wave functions.
\newblock Chem Phys Lett 2005;412:121--124.

\bibitem[{Bunge and Carb{\'o}-Dorca(2006)Bunge, Carlos F. and Carb{\'o}-Dorca,
  Ramon}]{Bunge_2006}
Bunge CF, Carb{\'o}-Dorca R.
\newblock Select-divide-and-conquer method for large-scale configuration
  interaction.
\newblock J Chem Phys 2006;125:014108.

\bibitem[{Bytautas and Ruedenberg(2009)Bytautas, Laimutis and Ruedenberg,
  Klaus}]{Bytautas_2009}
Bytautas L, Ruedenberg K.
\newblock A priori identification of configurational deadwood.
\newblock Chem Phys 2009;356:64--75.

\bibitem[{Giner et~al.(2013)Giner, Emmanuel and Scemama, Anthony and Caffarel,
  Michel}]{Giner_2013}
Giner E, Scemama A, Caffarel M.
\newblock Using perturbatively selected configuration interaction in quantum
  Monte Carlo calculations.
\newblock Can J Chem 2013;91:879--885.

\bibitem[{Caffarel et~al.(2014)Caffarel, Michel and Giner, Emmanuel and
  Scemama, Anthony and Ram{\'\i}rez-Sol{\'\i}s, Alejandro}]{Caffarel_2014}
Caffarel M, Giner E, Scemama A, Ram{\'\i}rez-Sol{\'\i}s A.
\newblock Spin Density Distribution in Open-Shell Transition Metal Systems: A
  Comparative Post-Hartree--Fock, Density Functional Theory, and Quantum Monte
  Carlo Study of the CuCl2Molecule.
\newblock J Chem Theory Comput 2014;10:5286--5296.

\bibitem[{Giner et~al.(2015)Emmanuel Giner and Anthony Scemama and Michel
  Caffarel}]{Giner_2015}
Giner E, Scemama A, Caffarel M.
\newblock Fixed-node diffusion Monte Carlo potential energy curve of the
  fluorine molecule F2 using selected configuration interaction trial
  wavefunctions.
\newblock J Chem Phys 2015;142:044115.

\bibitem[{Garniron et~al.(2017)Garniron, Yann and Scemama, Anthony and Loos,
  Pierre-Fran{\c c}ois and Caffarel, Michel}]{Garniron_2017b}
Garniron Y, Scemama A, Loos PF, Caffarel M.
\newblock Hybrid stochastic-deterministic calculation of the second-order
  perturbative contribution of multireference perturbation theory.
\newblock J Chem Phys 2017;147:034101.

\bibitem[{Caffarel et~al.(2016{\natexlab{a}})Caffarel, Michel and Applencourt,
  Thomas and Giner, Emmanuel and Scemama, Anthony}]{Caffarel_2016a}
Caffarel M, Applencourt T, Giner E, Scemama A.
\newblock {Communication: Toward an improved control of the fixed-node error in
  quantum Monte Carlo: The case of the water molecule}.
\newblock J Chem Phys 2016;144:151103.

\bibitem[{Caffarel et~al.(2016{\natexlab{b}})Michel Caffarel and Thomas
  Applencourt and Emmanuel Giner and Anthony Scemama}]{Caffarel_2016b}
Caffarel M, Applencourt T, Giner E, Scemama A.
\newblock 2.
\newblock In: Using CIPSI Nodes in Diffusion Monte Carlo; 2016. p. 15--46.

\bibitem[{Holmes et~al.(2016)Holmes, Adam A. and Tubman, Norm M. and Umrigar,
  C. J.}]{Holmes_2016}
Holmes AA, Tubman NM, Umrigar CJ.
\newblock Heat-Bath Configuration Interaction: An Efficient Selected
  Configuration Interaction Algorithm Inspired by Heat-Bath Sampling.
\newblock J Chem Theory Comput 2016;12:3674--3680.

\bibitem[{Sharma et~al.(2017)Sharma, Sandeep and Holmes, Adam A. and
  Jeanmairet, Guillaume and Alavi, Ali and Umrigar, C. J.}]{Sharma_2017}
Sharma S, Holmes AA, Jeanmairet G, Alavi A, Umrigar CJ.
\newblock Semistochastic Heat-Bath Configuration Interaction Method: Selected
  Configuration Interaction with Semistochastic Perturbation Theory.
\newblock J Chem Theory Comput 2017;13:1595--1604.

\bibitem[{Scemama et~al.(2018{\natexlab{a}})Scemama, Anthony and Garniron, Yann
  and Caffarel, Michel and Loos, Pierre-Fran{\c c}ois}]{Scemama_2018}
Scemama A, Garniron Y, Caffarel M, Loos PF.
\newblock {Deterministic Construction of Nodal Surfaces within Quantum Monte
  Carlo: The Case of FeS}.
\newblock J Chem Theory Comput 2018;14:1395--1402.

\bibitem[{Scemama et~al.(2018{\natexlab{b}})Scemama, Anthony and Benali, Anouar
  and Jacquemin, Denis and Caffarel, Michel and Loos, Pierre-Fran{\c
  c}ois}]{Scemama_2018b}
Scemama A, Benali A, Jacquemin D, Caffarel M, Loos PF.
\newblock {Excitation energies from diffusion Monte Carlo using selected
  configuration interaction nodes}.
\newblock J Chem Phys 2018;149:034108.

\bibitem[{Garniron et~al.(2018)Y. Garniron and A. Scemama and E. Giner and M.
  Caffarel and P. F. Loos}]{Garniron_2018}
Garniron Y, Scemama A, Giner E, Caffarel M, Loos PF.
\newblock Selected Configuration Interaction Dressed by Perturbation.
\newblock J Chem Phys 2018;149:064103.

\bibitem[{Evangelista(2014)Evangelista, Francesco A.}]{Evangelista_2014}
Evangelista FA.
\newblock Adaptive multiconfigurational wave functions.
\newblock J Chem Phys 2014;140:124114.

\bibitem[{Tubman et~al.(2016)Tubman, Norm M. and Lee, Joonho and Takeshita,
  Tyler Y. and {Head-Gordon}, Martin and Whaley, K. Birgitta}]{Tubman_2016}
Tubman NM, Lee J, Takeshita TY, {Head-Gordon} M, Whaley KB.
\newblock A Deterministic Alternative to the Full Configuration Interaction
  Quantum {{Monte Carlo}} Method.
\newblock J Chem Phys 2016;145:044112.

\bibitem[{Tubman et~al.(2020)Tubman, N. M. and Freeman, C. D. and Levine, D. S.
  and Hait, D. and Head-Gordon, M. and Whaley, K. B.}]{Tubman_2020}
Tubman NM, Freeman CD, Levine DS, Hait D, Head-Gordon M, Whaley KB.
\newblock Modern Approaches to Exact Diagonalization and Selected Configuration
  Interaction with the Adaptive Sampling CI Method.
\newblock J Chem Theory Comput 2020;16:2139.

\bibitem[{Schriber and Evangelista(2016)Schriber, Jeffrey B. and Evangelista,
  Francesco A.}]{Schriber_2016}
Schriber JB, Evangelista FA.
\newblock Communication: {An} adaptive configuration interaction approach for
  strongly correlated electrons with tunable accuracy.
\newblock J Chem Phys 2016;144:161106.

\bibitem[{Schriber and Evangelista(2017)Schriber, Jeffrey B. and Evangelista,
  Francesco A.}]{Schriber_2017}
Schriber JB, Evangelista FA.
\newblock {Adaptive Configuration Interaction for Computing Challenging
  Electronic Excited States with Tunable Accuracy}.
\newblock J Chem Theory Comput 2017;.

\bibitem[{Liu and Hoffmann(2016)Liu, Wenjian and Hoffmann, Mark R.}]{Liu_2016}
Liu W, Hoffmann MR.
\newblock {iCI: Iterative CI toward full CI}.
\newblock J Chem Theory Comput 2016;12:1169--1178.

\bibitem[{Per and Cleland(2017)Per, Manolo C. and Cleland, Deidre
  M.}]{Per_2017}
Per MC, Cleland DM.
\newblock Energy-based truncation of multi-determinant wavefunctions in quantum
  Monte Carlo.
\newblock J Chem Phys 2017;146:164101.

\bibitem[{Ohtsuka and Hasegawa(2017)Ohtsuka, Yuhki and Hasegawa,
  Jun-ya}]{Ohtsuka_2017}
Ohtsuka Y, Hasegawa Jy.
\newblock Selected configuration interaction method using sampled first-order
  corrections to wave functions.
\newblock J Chem Phys 2017;147:034102.

\bibitem[{Zimmerman(2017)Zimmerman, Paul M.}]{Zimmerman_2017}
Zimmerman PM.
\newblock Incremental full configuration interaction.
\newblock J Chem Phys 2017;146:104102.

\bibitem[{Coe(2018)J. P. Coe}]{Coe_2018}
Coe JP.
\newblock Machine Learning Configuration Interaction.
\newblock J Chem Theory Comput 2018;14:5739.

\bibitem[{Liu and Hoffmann(2014)Liu, W. and Hoffmann, M.R.}]{Liu_2014}
Liu W, Hoffmann MR.
\newblock SDS: the static--dynamic--static framework for strongly correlated
  electrons.
\newblock Theor Chem Acc 2014;133:1481.

\bibitem[{Lei et~al.(2017)Yibo Lei and Wenjian Liu and Mark R.
  Hoffmann}]{Lei_2017}
Lei Y, Liu W, Hoffmann MR.
\newblock Further development of SDSPT2 for strongly correlated electrons.
\newblock Mol Phys 2017;115:2696--2707.

\bibitem[{Zhang et~al.(2020)Zhang, Ning and Liu, Wenjian and Hoffmann, Mark
  R.}]{Zhang_2020}
Zhang N, Liu W, Hoffmann MR.
\newblock Iterative Configuration Interaction with Selection.
\newblock J Chem Theory Comput 2020;16:2296--2316.

\bibitem[{Garniron et~al.(2019)Y. Garniron and K. Gasperich and T. Applencourt
  and A. Benali and A. Fert{\'e} and J. Paquier and B. Pradines and R. Assaraf
  and P. Reinhardt and J. Toulouse and P. Barbaresco and N. Renon and G. David
  and J. P. Malrieu and M. V{\'e}ril and M. Caffarel and P. F. Loos and E.
  Giner and A. Scemama}]{Garniron_2019}
Garniron Y, Gasperich K, Applencourt T, Benali A, Fert{\'e} A, Paquier J,
  et~al.
\newblock Quantum Package 2.0: A Open-Source Determinant-Driven Suite Of
  Programs.
\newblock J Chem Theory Comput 2019;15:3591.

\bibitem[{Evangelisti et~al.(1983)Evangelisti, Stefano and Daudey, Jean-Pierre
  and Malrieu, Jean-Paul}]{Evangelisti_1983}
Evangelisti S, Daudey JP, Malrieu JP.
\newblock Convergence of an improved CIPSI algorithm.
\newblock Chem Phys 1983;75:91--102.

\bibitem[{Cimiraglia(1985)Cimiraglia, Renzo}]{Cimiraglia_1985}
Cimiraglia R.
\newblock Second order perturbation correction to {CI} energies by use of
  diagrammatic techniques: {An} improvement to the {CIPSI} algorithm.
\newblock J Chem Phys 1985;83:1746--1749.

\bibitem[{Cimiraglia and Persico(1987)Cimiraglia, Renzo and Persico,
  Maurizio}]{Cimiraglia_1987}
Cimiraglia R, Persico M.
\newblock Recent advances in multireference second order perturbation {CI}:
  {The} {CIPSI} method revisited.
\newblock J Comput Chem 1987;8:39--47.

\bibitem[{Illas et~al.(1988)Illas, F. and Rubio, J. and Ricart, J.
  M.}]{Illas_1988}
Illas F, Rubio J, Ricart JM.
\newblock Approximate natural orbitals and the convergence of a second order
  multireference many?body perturbation theory ({CIPSI}) algorithm.
\newblock J Chem Phys 1988;89:6376--6384.

\bibitem[{Povill et~al.(1992)Povill, A. and Rubio, J. and Illas,
  F.}]{Povill_1992}
Povill A, Rubio J, Illas F.
\newblock Treating large intermediate spaces in the {CIPSI} method through a
  direct selected {CI} algorithm.
\newblock Theor Chem Acc 1992;82:229--238.

\bibitem[{Garniron et~al.(2017)Garniron, Yann and Scemama, Anthony and Loos,
  Pierre-Fran{\c c}ois and Caffarel, Michel}]{Garniron_2017}
Garniron Y, Scemama A, Loos PF, Caffarel M.
\newblock Hybrid Stochastic-Deterministic Calculation of the Second-Order
  Perturbative Contribution of Multireference Perturbation Theory.
\newblock J Chem Phys 2017;147:034101.

\bibitem[{Scemama et~al.(2015)Scemama, Anthony and Giner, Emmanuel and
  Applencourt, Thomas and Caffarel, Michel}]{Scemama_2015}
Scemama A, Giner E, Applencourt T, Caffarel M, {QMC using very large
  configuration interaction-type expansions}; 2015.
\newblock Pacifichem, Advances in Quantum Monte Carlo.

\bibitem[{Scemama et~al.(2016)Scemama, Anthony and Applencourt, Thomas and
  Giner, Emmanuel and Caffarel, Michel}]{Scemama_2016}
Scemama A, Applencourt T, Giner E, Caffarel M.
\newblock Quantum Monte Carlo with very large multideterminant wavefunctions.
\newblock J Comput Chem 2016;37:1866--1875.

\bibitem[{Scemama et~al.(2019)A. Scemama and M. Caffarel and A. Benali and D.
  Jacquemin and P. F. Loos.}]{Scemama_2019}
Scemama A, Caffarel M, Benali A, Jacquemin D, Loos PF.
\newblock Influence of pseudopotentials on excitation energies from selected
  configuration interaction and diffusion Monte Carlo.
\newblock Res Chem 2019;1:100002.

\bibitem[{Dash et~al.(2018)M. Dash and S. Moroni and A. Scemama and C.
  Filippi}]{Dash_2018}
Dash M, Moroni S, Scemama A, Filippi C.
\newblock Perturbatively selected configuration-interaction wave functions for
  efficient geometry optimization in quantum Monte Carlo.
\newblock arXiv:180409610 2018;.

\bibitem[{Dash et~al.(2019)Dash, Monika and Feldt, Jonas and Moroni, Saverio
  and Scemama, Anthony and Filippi, Claudia}]{Dash_2019}
Dash M, Feldt J, Moroni S, Scemama A, Filippi C.
\newblock {Excited States with Selected Configuration Interaction-Quantum Monte
  Carlo: Chemically Accurate Excitation Energies and Geometries}.
\newblock J Chem Theory Comput 2019;15:4896--4906.

\bibitem[{Scemama et~al.(2020)Scemama,Anthony and Giner,Emmanuel and
  Benali,Anouar and Loos,Pierre-Fran{\c c}ois}]{Scemama_2020}
Scemama A, Giner E, Benali A, Loos PF.
\newblock Taming the fixed-node error in diffusion Monte Carlo via range
  separation.
\newblock J Chem Phys 2020;153:174107.

\bibitem[{Benali et~al.(2020)Anouar Benali and Kevin Gasperich and Kenneth D.
  Jordan and Thomas Applencourt and Ye Luo and M. Chandler Bennett and Jaron T.
  Krogel and Luke Shulenburger and Paul R. C. Kent and Pierre-Fran{\c c}ois
  Loos and Anthony Scemama and Michel Caffarel}]{Benali_2020}
Benali A, Gasperich K, Jordan KD, Applencourt T, Luo Y, Bennett MC, et~al.
\newblock Towards a Systematic Improvement of the Fixed-Node Approximation in
  Diffusion Monte Carlo for Solids.
\newblock J Chem Phys 2020;153:184111.

\bibitem[{Budz{\'a}k et~al.(2017)Budz{\'a}k, {\v S}. and Scalmani, G. and
  Jacquemin, D.}]{Budzak_2017}
Budz{\'a}k {\v S}, Scalmani G, Jacquemin D.
\newblock Accurate Excited-State Geometries: a CASPT2 and Coupled-Cluster
  Reference Database for Small Molecules.
\newblock J Chem Theory Comput 2017;13:6237--6252.

\bibitem[{Aidas et~al.(2014)Aidas, Kestutis and Angeli, Celestino and Bak, Keld
  L. and Bakken, Vebj{\o}rn and Bast, Radovan and Boman, Linus and
  Christiansen, Ove and Cimiraglia, Renzo and Coriani, Sonia and Dahle, P{\aa}l
  and Dalskov, Erik K. and Ekstr{\"o}m, Ulf and Enevoldsen, Thomas and Eriksen,
  Janus J. and Ettenhuber, Patrick and Fern{\'a}ndez, Berta and Ferrighi, Lara
  and Fliegl, Heike and Frediani, Luca and Hald, Kasper and Halkier, Asger and
  H{\"a}ttig, Christof and Heiberg, Hanne and Helgaker, Trygve and Hennum, Alf
  Christian and Hettema, Hinne and Hjerten{\ae}s, Eirik and H{\o}st, Stinne and
  H{\o}yvik, Ida-Marie and Iozzi, Maria Francesca and Jans{\'\i}k, Branislav
  and Jensen, Hans J{\o}rgen Aa. and Jonsson, Dan and J{\o}rgensen, Poul and
  Kauczor, Joanna and Kirpekar, Sheela and Kj{\ae}rgaard, Thomas and Klopper,
  Wim and Knecht, Stefan and Kobayashi, Rika and Koch, Henrik and Kongsted,
  Jacob and Krapp, Andreas and Kristensen, Kasper and Ligabue, Andrea and
  Lutn{\ae}s, Ola B. and Melo, Juan I. and Mikkelsen, Kurt V. and Myhre, Rolf
  H. and Neiss, Christian and Nielsen, Christian B. and Norman, Patrick and
  Olsen, Jeppe and Olsen, J{\'o}gvan Magnus H. and Osted, Anders and Packer,
  Martin J. and Pawlowski, Filip and Pedersen, Thomas B. and Provasi, Patricio
  F. and Reine, Simen and Rinkevicius, Zilvinas and Ruden, Torgeir A. and Ruud,
  Kenneth and Rybkin, Vladimir V. and Sa{\l}ek, Pawel and Samson, Claire C. M.
  and de Mer{\'a}s, Alfredo S{\'a}nchez and Saue, Trond and Sauer, Stephan P.
  A. and Schimmelpfennig, Bernd and Sneskov, Kristian and Steindal, Arnfinn H.
  and Sylvester-Hvid, Kristian O. and Taylor, Peter R. and Teale, Andrew M. and
  Tellgren, Erik I. and Tew, David P. and Thorvaldsen, Andreas J. and
  Th{\o}gersen, Lea and Vahtras, Olav and Watson, Mark A. and Wilson, David J.
  D. and Ziolkowski, Marcin and {\AA}gren, Hans}]{dalton}
Aidas K, Angeli C, Bak KL, Bakken V, Bast R, Boman L, et~al.
\newblock The Dalton Quantum Chemistry Program System.
\newblock WIREs Comput Mol Sci 2014;4:269--284.

\bibitem[{cfo(????)}]{cfour}
;.
\newblock CFOUR, Coupled-Cluster techniques for Computational Chemistry, a
  quantum-chemical program package by J.F. Stanton, J. Gauss, L. Cheng, M.E.
  Harding, D.A. Matthews, P.G. Szalay with contributions from A.A. Auer, R.J.
  Bartlett, U. Benedikt, C. Berger, D.E. Bernholdt, Y.J. Bomble, O.
  Christiansen, F. Engel, R. Faber, M. Heckert, O. Heun, M. Hilgenberg, C.
  Huber, T.-C. Jagau, D. Jonsson, J. Jus{\'e}lius, T. Kirsch, K. Klein, W.J.
  Lauderdale, F. Lipparini, T. Metzroth, L.A. M{\"u}ck, D.P. O'Neill, D.R.
  Price, E. Prochnow, C. Puzzarini, K. Ruud, F. Schiffmann, W. Schwalbach, C.
  Simmons, S. Stopkowicz, A. Tajti, J. V{\'a}zquez, F. Wang, J.D. Watts and the
  integral packages MOLECULE (J. Alml{\"o}f and P.R. Taylor), PROPS (P.R.
  Taylor), ABACUS (T. Helgaker, H.J. Aa. Jensen, P. J{\o}rgensen, and J.
  Olsen), and ECP routines by A. V. Mitin and C. van W{\"u}llen. For the
  current version, see http://www.cfour.de.

\bibitem[{Frisch et~al.(2016)M. J. Frisch and G. W. Trucks and H. B. Schlegel
  and G. E. Scuseria and M. A. Robb and J. R. Cheeseman and G. Scalmani and V.
  Barone and G. A. Petersson and H. Nakatsuji and X. Li and M. Caricato and A.
  V. Marenich and J. Bloino and B. G. Janesko and R. Gomperts and B. Mennucci
  and H. P. Hratchian and J. V. Ortiz and A. F. Izmaylov and J. L. Sonnenberg
  and D. Williams-Young and F. Ding and F. Lipparini and F. Egidi and J. Goings
  and B. Peng and A. Petrone and T. Henderson and D. Ranasinghe and V. G.
  Zakrzewski and J. Gao and N. Rega and G. Zheng and W. Liang and M. Hada and
  M. Ehara and K. Toyota and R. Fukuda and J. Hasegawa and M. Ishida and T.
  Nakajima and Y. Honda and O. Kitao and H. Nakai and T. Vreven and K.
  Throssell and Montgomery, {Jr.}, J. A. and J. E. Peralta and F. Ogliaro and
  M. J. Bearpark and J. J. Heyd and E. N. Brothers and K. N. Kudin and V. N.
  Staroverov and T. A. Keith and R. Kobayashi and J. Normand and K.
  Raghavachari and A. P. Rendell and J. C. Burant and S. S. Iyengar and J.
  Tomasi and M. Cossi and J. M. Millam and M. Klene and C. Adamo and R. Cammi
  and J. W. Ochterski and R. L. Martin and K. Morokuma and O. Farkas and J. B.
  Foresman and D. J. Fox}]{Gaussian16}
Frisch MJ, Trucks GW, Schlegel HB, Scuseria GE, Robb MA, Cheeseman JR, et~al.,
  Gaussian 16 {R}evision {A}.03; 2016.
\newblock Gaussian Inc. Wallingford CT.

\bibitem[{Binkley and Pople(1977)Binkley, J. Stephen and Pople, John
  A.}]{Binkley_1977a}
Binkley JS, Pople JA.
\newblock Self-consistent molecular orbital methods. XIX. Split-valence
  Gaussian-type basis sets for beryllium.
\newblock J Chem Phys 1977;66:879--880.

\bibitem[{Clark et~al.(1983)Clark, Timothy and Chandrasekhar, Jayaraman and
  Spitznagel, G{\"u}nther W. and Schleyer, Paul Von Ragu{\'e}}]{Clark_1983a}
Clark T, Chandrasekhar J, Spitznagel GW, Schleyer PVR.
\newblock Efficient diffuse function-augmented basis sets for anion
  calculations. III. The 3-21+G basis set for first-row elements, Li-F.
\newblock J Comput Chem 1983;4:294--301.

\bibitem[{Dill and Pople(1975)Dill, James D. and Pople, John A.}]{Dill_1975a}
Dill JD, Pople JA.
\newblock Self-consistent molecular orbital methods. XV. Extended Gaussian-type
  basis sets for lithium, beryllium, and boron.
\newblock J Chem Phys 1975;62:2921--2923.

\bibitem[{Ditchfield et~al.(1971)Ditchfield, R. and Hehre, W. J. and Pople, J.
  A.}]{Ditchfield_1971a}
Ditchfield R, Hehre WJ, Pople JA.
\newblock Self-Consistent Molecular-Orbital Methods. IX. An Extended
  Gaussian-Type Basis for Molecular-Orbital Studies of Organic Molecules.
\newblock J Chem Phys 1971;54:724--728.

\bibitem[{Francl et~al.(1982)Francl, Michelle M. and Pietro, William J. and
  Hehre, Warren J. and Binkley, J. Stephen and Gordon, Mark S. and DeFrees,
  Douglas J. and Pople, John A.}]{Francl_1982a}
Francl MM, Pietro WJ, Hehre WJ, Binkley JS, Gordon MS, DeFrees DJ, et~al.
\newblock Self-consistent molecular orbital methods. XXIII. A polarization-type
  basis set for second-row elements.
\newblock J Chem Phys 1982;77:3654--3665.

\bibitem[{Gordon et~al.(1982)Gordon, Mark S. and Binkley, J. Stephen and Pople,
  John A. and Pietro, William J. and Hehre, Warren J.}]{Gordon_1982a}
Gordon MS, Binkley JS, Pople JA, Pietro WJ, Hehre WJ.
\newblock Self-consistent molecular-orbital methods. 22. Small split-valence
  basis sets for second-row elements.
\newblock J Am Chem Soc 1982;104:2797--2803.

\bibitem[{Hehre et~al.(1972)Hehre, W. J. and Ditchfield, R. and Pople, J.
  A.}]{Hehre_1972a}
Hehre WJ, Ditchfield R, Pople JA.
\newblock Self-Consistent Molecular Orbital Methods. XII. Further Extensions of
  Gaussian-Type Basis Sets for Use in Molecular Orbital Studies of Organic
  Molecules.
\newblock J Chem Phys 1972;56:2257--2261.

\bibitem[{Dunning(1989)Dunning, Thom H.}]{Dunning_1989a}
Dunning TH.
\newblock Gaussian basis sets for use in correlated molecular calculations. I.
  The atoms boron through neon and hydrogen.
\newblock J Chem Phys 1989;90:1007--1023.

\bibitem[{Kendall et~al.(1992)Kendall, Rick A. and Dunning, Thom H. and
  Harrison, Robert J.}]{Kendall_1992a}
Kendall RA, Dunning TH, Harrison RJ.
\newblock Electron affinities of the first-row atoms revisited. Systematic
  basis sets and wave functions.
\newblock J Chem Phys 1992;96:6796--6806.

\bibitem[{Prascher et~al.(2011)Prascher, Brian P. and Woon, David E. and
  Peterson, Kirk A. and Dunning, Thom H. and Wilson, Angela
  K.}]{Prascher_2011a}
Prascher BP, Woon DE, Peterson KA, Dunning TH, Wilson AK.
\newblock Gaussian basis sets for use in correlated molecular calculations.
  VII. Valence, core-valence, and scalar relativistic basis sets for Li, Be,
  Na, and Mg.
\newblock Theor Chem Acc 2011;128:69--82.

\bibitem[{Woon and Dunning(1993)Woon, David E. and Dunning, Thom
  H.}]{Woon_1993a}
Woon DE, Dunning TH.
\newblock Gaussian basis sets for use in correlated molecular calculations.
  III. The atoms aluminum through argon.
\newblock J Chem Phys 1993;98:1358--1371.

\bibitem[{Woon and Dunning(1994)Woon, David E. and Dunning, Thom
  H.}]{Woon_1994a}
Woon DE, Dunning TH.
\newblock Gaussian basis sets for use in correlated molecular calculations. IV.
  Calculation of static electrical response properties.
\newblock J Chem Phys 1994;100:2975--2988.

\bibitem[{Feller(1996)Feller, David}]{Feller_1996a}
Feller D.
\newblock The role of databases in support of computational chemistry
  calculations.
\newblock J Comput Chem 1996;17:1571--1586.

\bibitem[{Pritchard et~al.(2019)Pritchard, Benjamin P. and Altarawy, Doaa and
  Didier, Brett and Gibsom, Tara D. and Windus, Theresa L.}]{Pritchard_2019a}
Pritchard BP, Altarawy D, Didier B, Gibsom TD, Windus TL.
\newblock A New Basis Set Exchange: An Open, Up-to-date Resource for the
  Molecular Sciences Community.
\newblock J Chem Inf Model 2019;59:4814--4820.

\bibitem[{Schuchardt et~al.(2007)Schuchardt, Karen L. and Didier, Brett T. and
  Elsethagen, Todd and Sun, Lisong and Gurumoorthi, Vidhya and Chase, Jared and
  Li, Jun and Windus, Theresa L.}]{Schuchardt_2007a}
Schuchardt KL, Didier BT, Elsethagen T, Sun L, Gurumoorthi V, Chase J, et~al.
\newblock Basis Set Exchange: A Community Database for Computational Sciences.
\newblock J Chem Inf Model 2007;47:1045--1052.

\bibitem[{Rolik et~al.(2013)Zolt{\'a}n Rolik and L{\'o}r{\'a}nt Szegedy and
  Istv{\'a}n Ladj{\'a}nszki and Bence Lad{\'o}czki and Mih{\'a}ly
  K{\'a}llay}]{Rolik_2013}
Rolik Z, Szegedy L, Ladj{\'a}nszki I, Lad{\'o}czki B, K{\'a}llay M.
\newblock An Efficient Linear-Scaling CCSD(T) Method Based on Local Natural
  Orbitals.
\newblock J Chem Phys 2013;139:094105.

\bibitem[{K{\'a}llay et~al.(2017)M. K{\'a}llay and Z. Rolik and J. Csontos and
  P. Nagy and G. Samu and D. Mester and J. Cs{\'o}ka and B. Szab{\'o} and I.
  Ladj{\'a}nszki and L. Szegedy and B. Lad{\'o}czki and K. Petrov and M. Farkas
  and P. D. Mezei and B. H{\'e}gely.}]{mrcc}
K{\'a}llay M, Rolik Z, Csontos J, Nagy P, Samu G, Mester D, et~al., MRCC,
  Quantum Chemical Program; 2017.

\bibitem[{Applencourt et~al.(2018)Thomas Applencourt and Kevin Gasperich and
  Anthony Scemama}]{Applencourt_2018}
Applencourt T, Gasperich K, Scemama A, Spin adaptation with determinant-based
  selected configuration interaction; 2018.

\bibitem[{Tur(????)}]{Turbomole}
TURBOMOLE V7.3 2018, a development of University of Karlsruhe and
  Forschungszentrum Karlsruhe GmbH, 1989-2007, TURBOMOLE GmbH, since 2007;
  available from {\tt http://www.turbomole.com} (accessed 13 June 2016).;.

\bibitem[{Head-Gordon et~al.(1994)M. Head-Gordon and R. J. Rico and M. Oumi and
  T. J. Lee}]{Head-Gordon_1994}
Head-Gordon M, Rico RJ, Oumi M, Lee TJ.
\newblock A Doubles Correction To Electronic Excited States From Configuration
  Interaction In The Space Of Single Substitutions.
\newblock Chem Phys Lett 1994;.

\bibitem[{Head-Gordon et~al.(1995)Head-Gordon, M. and Maurice, D. and Oumi,
  M.}]{Head-Gordon_1995}
Head-Gordon M, Maurice D, Oumi M.
\newblock A Perturbative Correction to Restricted Open-Shell
  Configuration-Interaction with Single Substitutions for Excited-States of
  Radicals.
\newblock Chem Phys Lett 1995;246:114--121.

\bibitem[{Krylov and Gill(2013)Krylov, Anna I. and Gill, Peter
  M.W.}]{Krylov_2013}
Krylov AI, Gill PMW.
\newblock Q-Chem: an Engine for Innovation.
\newblock WIREs Comput Mol Sci 2013;3:317--326.

\bibitem[{Stanton and Gauss(1995)Stanton,John F. and
  Gauss,J{\"u}rgen}]{Stanton_1995c}
Stanton JF, Gauss J.
\newblock Perturbative Treatment of the Similarity Transformed Hamiltonian in
  Equation-of-Motion Coupled-Cluster Approximations.
\newblock J Chem Phys 1995;103:1064--1076.

\bibitem[{Trofimov et~al.(2002)Trofimov, A. B. and Stelter, G. and Schirmer,
  J.}]{Trofimov_2002}
Trofimov AB, Stelter G, Schirmer J.
\newblock Electron Excitation Energies Using a Consistent Third-Order
  Propagator Approach: {{Comparison}} with Full Configuration Interaction and
  Coupled Cluster Results.
\newblock J Chem Phys 2002;117:6402--6410.

\bibitem[{Harbach et~al.(2014)Harbach, Philipp H. P. and Wormit, Michael and
  Dreuw, Andreas}]{Harbach_2014}
Harbach PHP, Wormit M, Dreuw A.
\newblock The Third-Order Algebraic Diagrammatic Construction Method
  ({{ADC}}(3)) for the Polarization Propagator for Closed-Shell Molecules:
  {{Efficient}} Implementation and Benchmarking.
\newblock J Chem Phys 2014;141:064113.

\bibitem[{Trofimov and Schirmer(1997)A.B. Trofimov and J.
  Schirmer}]{Trofimov_1997}
Trofimov AB, Schirmer J.
\newblock Polarization Propagator Study of Electronic Excitation in key
  Heterocyclic Molecules I. Pyrrole.
\newblock Chem Phys 1997;214:153--170.

\bibitem[{Christiansen et~al.(1996)Ove Christiansen and Henrik Koch and Poul
  J{\o}rgensen}]{Christiansen_1996b}
Christiansen O, Koch H, J{\o}rgensen P.
\newblock Perturbative Triple Excitation Corrections to Coupled Cluster Singles
  and Doubles Excitation Energies.
\newblock J Chem Phys 1996;105:1451--1459.

\bibitem[{Watts and Bartlett(1996)John D. Watts and Rodney J.
  Bartlett}]{Watts_1996b}
Watts JD, Bartlett RJ.
\newblock Iterative and Non-Iterative Triple Excitation Corrections in
  Coupled-Cluster Methods for Excited Electronic States: the EOM-CCSDT-3 and
  EOM-CCSD($\tilde{T}$) Methods.
\newblock Chem Phys Lett 1996;258:581--588.

\bibitem[{Prochnow et~al.(2010)Prochnow, Eric and Harding, Michael E. and
  Gauss, J{\"u}rgen}]{Prochnow_2010}
Prochnow E, Harding ME, Gauss J.
\newblock Parallel Calculation of CCSDT and Mk-MRCCSDT Energies.
\newblock J Chem Theory Comput 2010;6:2339--2347.

\bibitem[{Neese(2012)Frank Neese}]{Neese_2012}
Neese F.
\newblock The ORCA Program System.
\newblock WIREs Comput Mol Sci 2012;2:73--78.

\bibitem[{Nooijen and Bartlett(1997)Marcel Nooijen and Rodney J.
  Bartlett}]{Nooijen_1997}
Nooijen M, Bartlett RJ.
\newblock A New Method for Excited States: Similarity Transformed
  Equation-Of-Motion Coupled-Cluster Theory.
\newblock J Chem Phys 1997;106:6441--6448.

\bibitem[{Dutta et~al.(2018)Dutta, Achintya Kumar and Nooijen, Marcel and
  Neese, Frank and Izs{\'a}k, R{\'o}bert}]{Dutta_2018}
Dutta AK, Nooijen M, Neese F, Izs{\'a}k R.
\newblock Exploring the Accuracy of a Low Scaling Similarity Transformed
  Equation of Motion Method for Vertical Excitation Energies.
\newblock J Chem Theory Comput 2018;14:72--91.

\bibitem[{Krauter et~al.(2013)Krauter, Caroline M. and Pernpointner, Markus and
  Dreuw, Andreas}]{Krauter_2013}
Krauter CM, Pernpointner M, Dreuw A.
\newblock Application of the Scaled-Opposite-Spin Approximation to Algebraic
  Diagrammatic Construction Schemes of Second Order.
\newblock J Chem Phys 2013;138:044107.

\bibitem[{Hellweg et~al.(2008)Hellweg, A. and Gr\"un, S. A. and H\"attig,
  C.}]{Hellweg_2008}
Hellweg A, Gr\"un SA, H\"attig C.
\newblock Benchmarking the Performance of Spin-Component Scaled CC2 in Ground
  and Electronically Excited States.
\newblock Phys Chem Chem Phys 2008;10:4119--4127.

\bibitem[{Parrish et~al.(2017)Parrish, Robert M. and Burns, Lori A. and Smith,
  Daniel G. A. and Simmonett, Andrew C. and DePrince, A. Eugene and Hohenstein,
  Edward G. and Bozkaya, U{\u g}ur and Sokolov, Alexander Yu. and Di Remigio,
  Roberto and Richard, Ryan M. and Gonthier, J{\'e}r{\^o}me F. and James,
  Andrew M. and McAlexander, Harley R. and Kumar, Ashutosh and Saitow, Masaaki
  and Wang, Xiao and Pritchard, Benjamin P. and Verma, Prakash and Schaefer,
  Henry F. and Patkowski, Konrad and King, Rollin A. and Valeev, Edward F. and
  Evangelista, Francesco A. and Turney, Justin M. and Crawford, T. Daniel and
  Sherrill, C. David}]{Psi4}
Parrish RM, Burns LA, Smith DGA, Simmonett AC, DePrince AE, Hohenstein EG,
  et~al.
\newblock Psi4 1.1: An Open-Source Electronic Structure Program Emphasizing
  Automation, Advanced Libraries, and Interoperability.
\newblock J Chem Theory Comput 2017;13:3185--3197.
\newblock PMID: 28489372.

\bibitem[{Pito{\v n}{\'a}k et~al.(2009)Pito{\v n}{\'a}k, Michal and
  Neogr{\'a}dy, Pavel and {\v C}ern{\'y}, Ji{\v r}{\'\i} and Grimme, Stefan and
  Hobza, Pavel}]{Pitonak_2009}
Pito{\v n}{\'a}k M, Neogr{\'a}dy P, {\v C}ern{\'y} J, Grimme S, Hobza P.
\newblock Scaled MP3 Non-Covalent Interaction Energies Agree Closely with
  Accurate CCSD(T) Benchmark Data.
\newblock ChemPhysChem 2009;10:282--289.

\bibitem[{Loos and Jacquemin(2020)Loos, Pierre-Francois and Jacquemin,
  Denis}]{Loos_2020d}
Loos PF, Jacquemin D.
\newblock Is ADC(3) as Accurate as CC3 for Valence and Rydberg Transition
  Energies?
\newblock J Phys Chem Lett 2020;11:974--980.

\bibitem[{Werner et~al.(2011)Werner, Hans-Joachim and Knowles, Peter J. and
  Knizia, Gerald and Manby, Frederick R. and Sch{\"u}tz, Martin}]{molpro}
Werner HJ, Knowles PJ, Knizia G, Manby FR, Sch{\"u}tz M.
\newblock Molpro: a general-purpose quantum chemistry program package.
\newblock WIREs Comput Mol Sci 2011;2:242--253.

\bibitem[{Angeli et~al.(2001{\natexlab{a}})Angeli, Celestino and Cimiraglia,
  Renzo and Malrieu, Jean-Paul}]{Angeli_2001a}
Angeli C, Cimiraglia R, Malrieu JP.
\newblock N-Electron Valence State Perturbation Theory: A Fast Implementation
  of the Strongly Contracted Variant.
\newblock Chem Phys Lett 2001;350:297--305.

\bibitem[{Angeli et~al.(2001{\natexlab{b}})Angeli, C. and Cimiraglia, R. and
  Evangelisti, S. and Leininger, T. and Malrieu, J.-P.}]{Angeli_2001b}
Angeli C, Cimiraglia R, Evangelisti S, Leininger T, Malrieu JP.
\newblock Introduction of {\emph{n}} -Electron Valence States for
  Multireference Perturbation Theory.
\newblock J Chem Phys 2001;114:10252--10264.

\bibitem[{Angeli et~al.(2002)Angeli, Celestino and Cimiraglia, Renzo and
  Malrieu, Jean-Paul}]{Angeli_2002}
Angeli C, Cimiraglia R, Malrieu JP.
\newblock {\emph{N}} -Electron Valence State Perturbation Theory: {{A}}
  Spinless Formulation and an Efficient Implementation of the Strongly
  Contracted and of the Partially Contracted Variants.
\newblock J Chem Phys 2002;117:9138--9153.

\bibitem[{Finley et~al.(1998)James Finley and Per-{\AA}ke Malmqvist and
  Bj{\"o}rn O. Roos and Luis Serrano-Andr{\'e}s}]{Finley_1998}
Finley J, Malmqvist P{\AA}, Roos BO, Serrano-Andr{\'e}s L.
\newblock The Multi-State CASPT2 Method.
\newblock Chem Phys Lett 1998;288:299--306.

\bibitem[{Shiozaki et~al.(2011)Shiozaki, Toru and Gy{\H o}rffy, Werner and
  Celani, Paolo and Werner, Hans-Joachim}]{Shiozaki_2011}
Shiozaki T, Gy{\H o}rffy W, Celani P, Werner HJ.
\newblock Communication: {{Extended}} Multi-State Complete Active Space
  Second-Order Perturbation Theory: {{Energy}} and Nuclear Gradients.
\newblock J Chem Phys 2011;135:081106.

\bibitem[{Svensson et~al.(1996{\natexlab{a}})Svensson, Mats and Humbel,
  St{\'e}phane and Froese, Robert D. J. and Matsubara, Toshiaki and Sieber,
  Stefan and Morokuma, Keiji}]{Svensson_1996a}
Svensson M, Humbel S, Froese RDJ, Matsubara T, Sieber S, Morokuma K.
\newblock ONIOM: A Multilayered Integrated MO + MM Method for Geometry
  Optimizations and Single Point Energy Predictions. A Test for Diels-Alder
  Reactions and Pt(P(t-Bu)3)2 + H2 Oxidative Addition.
\newblock J Phys Chem 1996;100:19357--19363.

\bibitem[{Svensson et~al.(1996{\natexlab{b}})Svensson,Mats and
  Humbel,St{\'e}phane and Morokuma,Keiji}]{Svensson_1996b}
Svensson M, Humbel S, Morokuma K.
\newblock Energetics using the single point IMOMO (integrated molecular
  orbital+molecular orbital) calculations: Choices of computational levels and
  model system.
\newblock J Chem Phys 1996;105:3654--3661.

\bibitem[{Caricato et~al.(2010)Caricato, M. and Trucks, G. W. and Frisch, M. J.
  and Wiberg, K. B.}]{Caricato_2010}
Caricato M, Trucks GW, Frisch MJ, Wiberg KB.
\newblock Electronic Transition Energies: A Study of the Performance of a Large
  Range of Single Reference Density Functional and Wave Function Methods on
  Valence and Rydberg States Compared to Experiment.
\newblock J Chem Theory Comput 2010;6:370--383.

\bibitem[{Watson et~al.(2013)Watson, Thomas J. and Lotrich, Victor F. and
  Szalay, Peter G. and Perera, Ajith and Bartlett, Rodney J.}]{Watson_2013}
Watson TJ, Lotrich VF, Szalay PG, Perera A, Bartlett RJ.
\newblock Benchmarking for Perturbative Triple-Excitations in EE-EOM-CC
  Methods.
\newblock J Phys Chem A 2013;117:2569--2579.

\bibitem[{K{\'a}nn{\'a}r and Szalay(2014)K{\'a}nn{\'a}r, D{\'a}niel and Szalay,
  P{\'e}ter G.}]{Kannar_2014}
K{\'a}nn{\'a}r D, Szalay PG.
\newblock Benchmarking Coupled Cluster Methods on Valence Singlet Excited
  States.
\newblock J Chem Theory Comput 2014;10:3757--3765.

\bibitem[{K{\'a}nn{\'a}r et~al.(2017)K{\'a}nn{\'a}r, D{\'a}niel and Tajti,
  Attila and Szalay, P{\'e}ter G.}]{Kannar_2017}
K{\'a}nn{\'a}r D, Tajti A, Szalay PG.
\newblock Accuracy of Coupled Cluster Excitation Energies in Diffuse Basis
  Sets.
\newblock J Chem Theory Comput 2017;13:202--209.

\bibitem[{Hodecker et~al.(2019)Hodecker,Manuel and Rehn,Dirk R. and
  Dreuw,Andreas and H{\"o}fener,Sebastian}]{Hodecker_2019}
Hodecker M, Rehn DR, Dreuw A, H{\"o}fener S.
\newblock Similarities and Differences of the Lagrange Formalism and the
  Intermediate State Representation in the Treatment of Molecular Properties.
\newblock J Chem Phys 2019;150:164125.

\bibitem[{Eriksen and Gauss(2020)Eriksen,Janus J. and
  Gauss,J{\"u}rgen}]{Eriksen_2020b}
Eriksen JJ, Gauss J.
\newblock Ground and excited state first-order properties in many-body expanded
  full configuration interaction theory.
\newblock J Chem Phys 2020;153:154107.

\bibitem[{Chrayteh et~al.(in press)A. Chrayteh and A. Blondel and P. F. Loos
  and D. Jacquemin}]{Chrayteh_2021}
Chrayteh A, Blondel A, Loos PF, Jacquemin D.
\newblock A mountaineering strategy to excited states: highly-accurate
  oscillator strengths and dipole moments of small molecules.
\newblock J Chem Theory Comput in press;.

\bibitem[{Sarkar et~al.(submitted)R. Sarkar and M. Boggio-Pasqua and P. F. Loos
  and D. Jacquemin}]{Sarkar_2021}
Sarkar R, Boggio-Pasqua M, Loos PF, Jacquemin D.
\newblock Benchmark of TD-DFT and Wavefunction Methods for Oscillator Strengths
  and Excited-State Dipoles.
\newblock J Chem Theory Comput submitted;.

\bibitem[{Jacquemin(2018)Jacquemin, Denis}]{Jacquemin_2018}
Jacquemin D.
\newblock What is the Key for Accurate Absorption and Emission Calculations ?
  Energy or Geometry ?
\newblock J Chem Theory Comput 2018;14:1534--1543.

\bibitem[{Gould(2018)Gould, Tim}]{Gould_2018b}
Gould T.
\newblock `Diet GMTKN55' offers accelerated benchmarking through a
  representative subset approach.
\newblock Phys Chem Chem Phys 2018;20:27735--27739.

\end{thebibliography}

%%%%%%%%%%%%%%%%%%%%%%%%%%%%%%%%

\newpage

\graphicalabstract{TOC}{QUEST: a dataset of highly-accurate excitation energies.}

\end{document}